\documentclass[11pt]{article}
\usepackage[T1]{fontenc}
\usepackage{lmodern}
\usepackage[letterpaper,margin=1in,headheight=14pt]{geometry}
\usepackage{microtype}
\usepackage[numbers,sort&compress]{natbib}
\setcitestyle{numbers,square,citesep={,}}
\usepackage{fancyhdr}
\usepackage[font=small,labelfont=bf]{caption}

\usepackage{amsmath,amssymb,amsthm}

\usepackage{amsmath,amsfonts,bm}

\def\eqref#1{equation~\ref{#1}}

\def\1{\bm{1}}

\DeclareMathAlphabet{\mathsfit}{\encodingdefault}{\sfdefault}{m}{sl}
\SetMathAlphabet{\mathsfit}{bold}{\encodingdefault}{\sfdefault}{bx}{n}

\usepackage{graphicx}
\usepackage{booktabs}
\usepackage{multirow}
\usepackage{xcolor}
\usepackage{algorithm}
\usepackage{algorithmic}
\usepackage{enumitem}
\usepackage{hyperref}
\usepackage{xurl}
\usepackage{url}
\usepackage[capitalize,nameinlink]{cleveref}
\graphicspath{{figures/}}

\usepackage{tcolorbox}
\newtcolorbox{propbox}{colback=blue!5!white, colframe=blue!40!black, boxrule=0.5pt, arc=3pt, left=1pt, right=1pt, top=2pt, bottom=2pt}

\newtheorem{theorem}{Theorem}
\newtheorem{proposition}[theorem]{Proposition}

\theoremstyle{definition}

\theoremstyle{plain}
\crefname{thm}{theorem}{theorems}\Crefname{thm}{Theorem}{Theorems}
\crefname{lem}{lemma}{lemmas}\Crefname{lem}{Lemma}{Lemmas}
\crefname{prop}{proposition}{propositions}\Crefname{prop}{Proposition}{Propositions}
\crefname{cor}{corollary}{corollaries}\Crefname{cor}{Corollary}{Corollaries}
\crefname{defn}{definition}{definitions}\Crefname{defn}{Definition}{Definitions}
\theoremstyle{remark}

\crefname{rem}{remark}{remarks}\Crefname{rem}{Remark}{Remarks}
\crefname{assum}{assumption}{assumptions}\Crefname{assum}{Assumption}{Assumptions}

\newcommand{\dd}{\mathrm{d}}

\title{VESPER: Online Learning of Free Energies\\ with a Multi-Tempered Variational Objective}

\author{\textbf{Jay L.~Kaplan\textsuperscript{1}, Gustavo R.~P\'erez-Lemus\textsuperscript{2}, Mengjian Hua\textsuperscript{1},}\\
\textbf{Juan J.~de Pablo\textsuperscript{1,2,3,*,\textdagger} \& Eric Vanden-Eijnden\textsuperscript{1,4,*,\textdagger}}\\[0.5em]
\textsuperscript{\textbf{1}}Courant Institute School of Mathematics, Computing, and Data Science,\\
New York University, New York, NY 10012, USA\\[0.3em]
\textsuperscript{\textbf{2}}Department of Chemical and Biomolecular Engineering, Tandon School of Engineering, \\New York University, Brooklyn, NY 11201, USA\\[0.3em]
\textsuperscript{\textbf{3}}Department of Physics, New York University, New York, NY 10012, USA\\[0.5em]
\textsuperscript{\textbf{4}}Machine Learning Lab, Capital Fund Management\\
\textsuperscript{*}Corresponding authors: \protect\href{mailto:jjd8110@nyu.edu}{jjd8110@nyu.edu}, \protect\href{mailto:eve2@cims.nyu.edu}{eve2@cims.nyu.edu}\\
\textsuperscript{\textdagger}Equal advising.}

\hypersetup{colorlinks=true,linkcolor=blue!55!black,citecolor=blue!55!black,urlcolor=blue!55!black}
\renewcommand{\lhead}[1]{\fancyhead[L]{\small\itshape #1}}

\makeatletter
\renewcommand{\@maketitle}{%
  \newpage
  \null
  \vskip 1em%
  \begin{center}%
    {\LARGE\bfseries \@title \par}%
    \vskip 1.5em%
    {\small\lineskip .5em%
      \ignorespaces\@author\par}%
  \end{center}%
  \par
  \vskip 1.5em}
\makeatother

\begin{document}

\maketitle
\lhead{Preprint}

\begin{abstract}
Computing molecular free-energy landscapes over collective variables (CVs) is central to understanding molecular mechanisms across physics, chemistry, and biology, but poses a difficult learning problem without training data available \textit{a priori}. The microscopic potential is known, yet samples of the marginal CV distribution must be generated by molecular dynamics (MD), whose trajectories can remain trapped in metastable states and rarely cross the free-energy barriers governing conformational changes or chemical reactions. We introduce VESPER, which couples the generation of these samples to neural-network-driven free-energy learning: MD supplies configurations for learning, while the evolving model biases the MD to accelerate exploration and generate further training data. VESPER minimizes a single variational objective that is convex in the free energy and can combine several tempering levels, each using a distinct temperature-boosting factor to flatten the effective free-energy landscape and promote barrier crossing. Every level has the same physical free energy as its minimizer, allowing data from all levels to train one shared model without reweighting. Learning is driven by the difference in mean parameter gradients between biased MD configurations and auxiliary walkers that explore the flattened model directly in CV space. Replica exchange between levels can further accelerate exploration. We prove conditional convergence of the coupled overdamped population dynamics without replica exchange under log-Sobolev and Polyak--\L{}ojasiewicz assumptions. VESPER matches or improves on established enhanced-sampling methods on alanine dipeptide and yields self-consistent free-energy reconstructions from three to six CVs on alanine tetrapeptide. For methane dehydrogenation on Ni(511) with a machine-learned interatomic potential, it recovers two-CV Boltzmann marginals and a reactant-to-saddle free-energy difference along a three-CV path, both qualitatively consistent with the lower-dimensional reference. It also samples RMSD-defined folded and unfolded regions of chignolin while biasing seventeen CVs.

\end{abstract}

\section{Introduction}\label{sec:intro}
Molecular dynamics (MD) is a central tool for studying the structure and dynamics of molecular systems in physics, chemistry, and biology. A recurring goal is to compute the free-energy landscape over a set of informative collective variables (CVs), defined as the negative logarithm of their marginal density under the equilibrium distribution, multiplied by the thermal energy $k_BT$. This landscape quantifies the relative stability of metastable states and the barriers separating them. Given samples from this marginal, learning it would be a standard estimation problem. In MD, however, the microscopic potential is known, but the relevant configurations must be generated by simulation. Brute-force MD is often computationally prohibitive: barrier-crossing transitions between metastable states can occur on microsecond-to-millisecond timescales, whereas integration requires femtosecond timesteps \citep{heninEnhancedSamplingMethods2022}. Free-energy learning is therefore coupled to a sampling problem, and the model must accelerate the simulation while it is itself being learned.

We introduce VESPER (Variational Extended Space Parallel Ensemble Regression), which learns a neural free energy $G_\theta$ while using it to bias the simulations that produce its own training data. VESPER minimizes a single variational objective: a positively weighted sum of components, one per tempering level, each using a distinct temperature-boosting factor to flatten the effective free-energy landscape. Each component is convex in the free energy and is minimized by the physical free energy at every such level. The sum therefore has the same minimizer up to an additive constant, and every level contributes its gradient to one model without reweighting to the physical ensemble. VESPER couples biased microscopic dynamics, persistent auxiliary CV walkers, and free-energy learning through one shared model. Through the CV map $s(x)$, the same $G_\theta$ sets both the MD bias and the tempered auxiliary potential, with no separate free-energy recursion. We also establish conditional convergence of the coupled sampling and learning dynamics. These contributions build on established variational connections: individual components extend the pseudo-spherical family to negative exponents and admit R\'enyi variational forms \citep{yu2021pseudospherical,fujisawa2008robust,birrell2021renyi}, and their gradients recover the instantaneous update of well-tempered variationally enhanced sampling (VES), up to scale, when its target is held fixed during differentiation \citep{Valsson_2014,valsson2015welltempered,bonatiNeuralNetworksbasedVariationally2019a}.

The gradient of each component is the difference between two averages of $\nabla_\theta G_\theta$: one over MD configurations evolving under the microscopic potential plus a bias built from $G_\theta$, the other over persistent auxiliary walkers evolving directly in CV space under a tempered $G_\theta$. At the optimum both populations sample the same flattened distribution, and the walkers interact with the molecular system only through the shared parameters, so no adiabatic limit is needed to identify the free energy, unlike in temperature-accelerated MD and driven adiabatic free-energy dynamics \citep{maraglianoTemperatureAcceleratedMethod2006a,abramsEfficientDirectGeneration2008a}. Replica exchange can additionally carry configurations discovered under strongly flattened biases down the ladder, facilitating barrier crossing along the CVs.

Our main contributions are:
\begin{itemize}[leftmargin=*]

\item \textbf{A single objective with a common minimizer.}
We establish that the losses associated with all tempering levels are convex in the free-energy function and share the physical free energy as their minimizer, up to an additive constant. Their positively weighted sum therefore defines one objective for one model, with every level contributing its native gradient without reweighting to the physical ensemble (\cref{thm:ladder}).

\item \textbf{Online learning from coupled sampling populations.}
Biased MD and persistent auxiliary CV walkers generate the configurations needed for learning as the model evolves, without a pre-existing dataset or an adiabatic limit between microscopic and auxiliary coordinates. We prove conditional convergence of the coupled overdamped population dynamics without replica exchange under log-Sobolev and Polyak--\L{}ojasiewicz assumptions, controlling both objective error and sampling lag (\cref{app:convergence}).

\item \textbf{Replica exchange on the learned free energy.}
The Metropolis acceptance probability for exchange between adjacent levels depends only on learned free-energy differences in lower-dimensional CV space, avoiding additional evaluations of the microscopic potential. At a fixed model, the exchange kernel preserves the product of the sampling laws across levels and therefore leaves the population gradient unchanged.

\item \textbf{Controlled tests of tempering and replica exchange.}
On the M\"uller--Brown potential, the tempering ladder balances basin accuracy, measured by KL divergence, with landscape reconstruction, measured by RMSE. On four-well landscapes spanning two to one hundred dimensions, a fixed ladder remains competitive without dimension-specific selection of $\alpha$, with replica exchange improving accuracy and convergence in the most challenging high-dimensional case (\cref{sec:toy-results}).

\item \textbf{Molecular free-energy reconstruction and high-dimensional exploration.}
On alanine dipeptide, VESPER achieves competitive reconstruction accuracy and convergence relative to the tested enhanced-sampling baselines. On alanine tetrapeptide, agreement between Boltzmann marginals learned with three and six CVs demonstrates dimensional self-consistency. For methane dehydrogenation on Ni(511) with a machine-learned interatomic potential, VESPER learns a three-CV landscape whose two-CV Boltzmann marginals and reactant-to-saddle free-energy difference along a three-CV path are consistent with the lower-dimensional reference. On chignolin, biasing seventeen dihedral CVs produces repeated RMSD-defined unfolding and refolding transitions, demonstrating effective exploration of a high-dimensional molecular CV space (\cref{sec:experiments}).

\end{itemize}

\crefalias{siproposition}{proposition}
\crefalias{sitheorem}{theorem}
\crefalias{sicorollary}{corollary}

\section{Methodology}
\label{sec:method}

\subsection{Collective variables and free energy}
\label{sec:setting}

Let $x\in\mathcal X$ denote a molecular configuration, $U(x)$ its microscopic potential energy, and $\varrho(x)=Z^{-1}e^{-\beta U(x)}$ its equilibrium density at inverse temperature $\beta=(k_BT)^{-1}$. A collective-variable map $s:\mathcal X\to\mathcal S$ reduces the $d$-dimensional configuration space to a $D$-dimensional CV domain, typically with $D\ll d$. The equilibrium CV density and its associated free energy are
\begin{equation}
\label{eq:physical-free-energy}
p(s)=\int_{\mathcal X}\delta(s-s(x))\varrho(x)\,\dd x,
\qquad
G_\star(s)=-\beta^{-1}\log p(s).
\end{equation}
The microscopic potential supplies the forces for simulation, whereas the marginal $p(s)$ must be inferred from configurations generated by the simulation. VESPER represents $G_\star$ by a differentiable model $G_\theta$ and uses it to bias the molecular dynamics that generate its training configurations.

\begin{figure*}[t]
\centering
  \includegraphics[width=\linewidth]{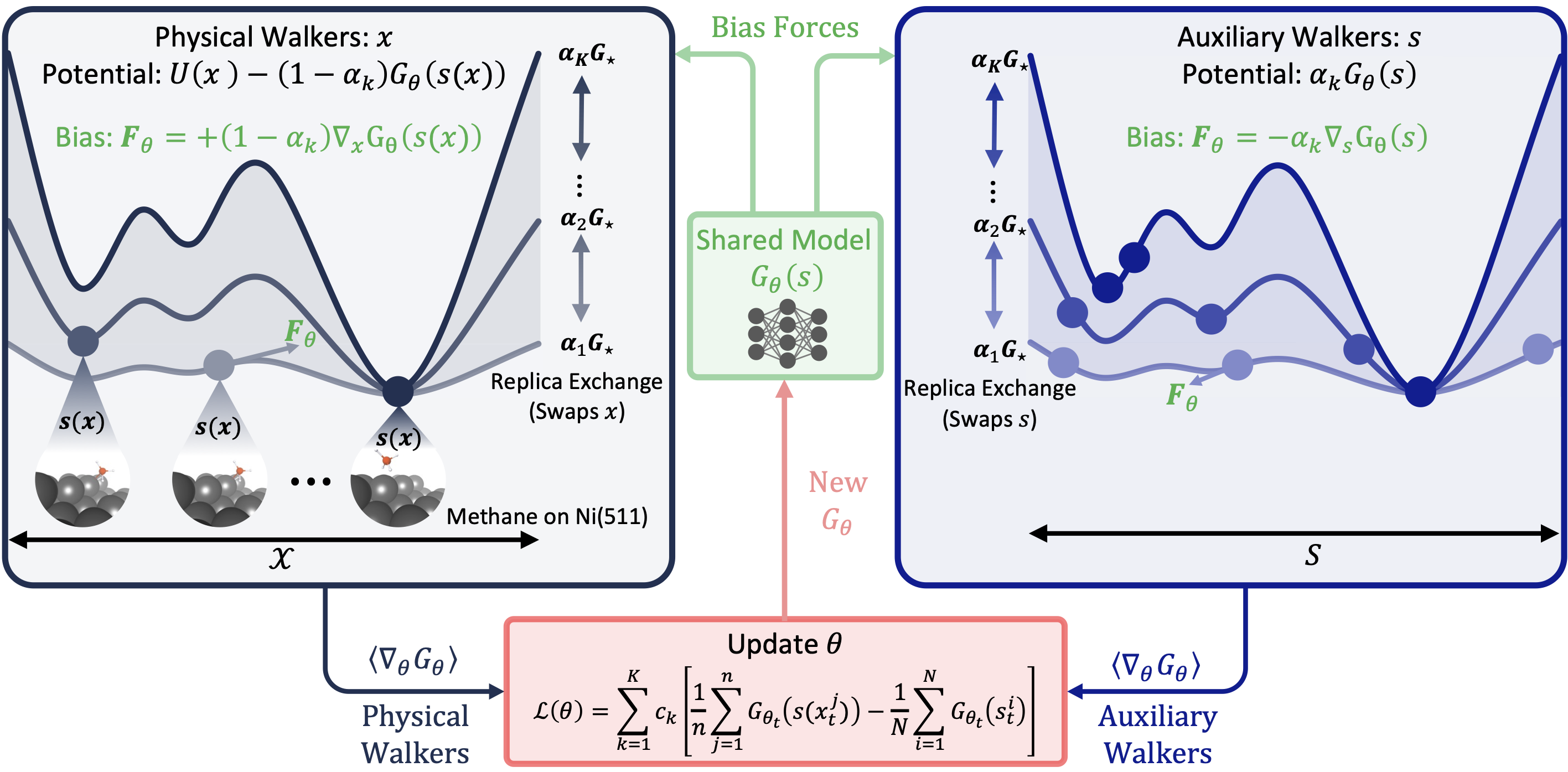}%
\caption{VESPER couples sampling and learning. Each tempering level $k$ carries two walker populations, and all levels share one model $G_\theta$. Left: physical walkers evolve in configuration space under $U(x)-(1-\alpha_k)G_\theta(s(x))$ and are mapped to CV space by $s(x)$. Replica exchange swaps configurations between adjacent levels. Right: auxiliary walkers evolve directly in CV space under $\alpha_kG_\theta(s)$. The two populations supply the averages in \eqref{eq:ladder-gradient}, whose $c_k$-weighted difference updates $\theta$. The updated model then sets the forces on both populations. At convergence, both populations at level $k$ sample the same flattened landscape $\alpha_kG_\star$. During learning, the instantaneous target for the physical CVs is $Q_{G_\theta,k}\propto p\,e^{(\beta-\hat\beta_k)G_\theta}$.}
\label{fig:diagram}
\end{figure*}

\subsection{The variational objective}
\label{sec:objective}

Fix an auxiliary inverse temperature $0<\hat\beta<\beta$ and set $\alpha=\hat\beta/\beta\in(0,1)$. We define the functional
\begin{equation}
\label{eq:objective}
\mathcal L_{\hat\beta}[G]
=
\frac{1}{\hat\beta}
\log
\int_{\mathcal S}e^{-\hat\beta G(s)}\,\dd s
+
\frac{1}{\beta-\hat\beta}
\log
\int_{\mathcal X}e^{(\beta-\hat\beta)G(s(x))}\varrho(x)\,\dd x.
\end{equation}

The two terms are the log-partition functions of a tempered version of the model on $\mathcal S$ and of the equilibrium density tilted by the model. They define the auxiliary CV law
\begin{equation}
\label{eq:single-level-laws}
P_{G,\hat\beta}(s)
=
\frac{e^{-\hat\beta G(s)}}{A_G}, \qquad A_G=\int_{\mathcal S}e^{-\hat\beta G(s)}\,\dd s
\end{equation}
and the biased microscopic law together with its CV marginal,
\begin{equation}
\label{eq:physical-pair}
\nu_{G,\hat\beta}(x)
=
\frac{e^{(\beta-\hat\beta)G(s(x))}\varrho(x)}{B_G},
\qquad
Q_{G,\hat\beta}(s)
=
\frac{p(s)e^{(\beta-\hat\beta)G(s)}}{B_G},
\qquad
B_G = \int_{\mathcal X}e^{(\beta-\hat\beta)G(s(x))}\varrho(x)\,\dd x.
\end{equation}

The first variation of \eqref{eq:objective} in a direction $h$ is $\mathbb E_{Q_{G,\hat\beta}}h-\mathbb E_{P_{G,\hat\beta}}h$, so $G$ is stationary exactly when the two CV laws coincide:
\begin{equation}
\frac{e^{-\hat\beta G}}{A_G}=\frac{p\,e^{(\beta-\hat\beta)G}}{B_G}
\iff
e^{-\beta G}\propto p
\iff
G=G_\star+C.
\end{equation}
The tempering exponent cancels: for every $\alpha$, the stationary point is the physical free energy. The second variation, $\hat\beta\,\mathrm{Var}_{P_{G,\hat\beta}}(h)+(\beta-\hat\beta)\,\mathrm{Var}_{Q_{G,\hat\beta}}(h)$, is nonnegative, so $\mathcal L_{\hat\beta}$ is convex and its minimizer set is $\{G_\star+C:C\in\mathbb R\}$ (\cref{thm:consistency}). At the minimizer,
\begin{equation}
\label{eq:common-tempered-law}
P_{G_\star,\hat\beta}
=
Q_{G_\star,\hat\beta}
=
\pi_\alpha,
\qquad
\pi_\alpha\propto e^{-\alpha\beta G_\star}\propto p^\alpha,
\end{equation}
so both populations explore the same flattened landscape, with barriers reduced by the factor $\alpha$. Near the minimizer,
$\mathcal L_{\hat\beta}[G]-\mathcal L_{\hat\beta}[G_\star]\approx\tfrac{\beta}{2}\mathrm{Var}_{\pi_\alpha}(G-G_\star)$:
$\alpha$ does not change the target, only where errors are weighted, with smaller $\alpha$ giving more weight to high free-energy regions.

For a parametric model $G_\theta$, the chain rule gives
\begin{equation}
\label{eq:physical-gradient}
\nabla_\theta\mathcal L_{\hat\beta}[G_\theta]
=
\mathbb E_{\nu_{G_\theta,\hat\beta}}
\!\left[\nabla_\theta G_\theta(s(x))\right]
-
\mathbb E_{P_{G_\theta,\hat\beta}}
\!\left[\nabla_\theta G_\theta(s)\right].
\end{equation}
Neither expectation requires knowing $p$. The first is over configurations sampled by MD biased with $G_\theta$, and the second over walkers sampling the tempered model in CV space (\cref{sec:algorithm}).

\subsection{The combined objective}
\label{sec:ladder}
Choose $0\leq\hat\beta_1<\cdots<\hat\beta_K\leq\beta$, set $\alpha_k=\hat\beta_k/\beta$, and write $P_{G,k}$, $\nu_{G,k}$, $Q_{G,k}$ for the laws of level $k$. The component $\mathcal L_{\hat\beta_k}$ is given by \eqref{eq:objective} for $0<\hat\beta_k<\beta$. At $\hat\beta_k=\beta$ and, on compact $\mathcal S$, at $\hat\beta_k=0$, define the endpoint objectives directly under the endpoint conditions in \cref{app:proof-kl}:
\begin{equation}
\label{eq:VESPER_objective:lim}
\mathcal L_\beta[G]
=
\frac{1}{\beta}\log\int_{\mathcal S}e^{-\beta G(s)}\,\dd s
+
\mathbb E_p[G],
\qquad
\mathcal L_0[G]
=
-\mathbb E_{u}[G]
+
\frac{1}{\beta}\log\mathbb E_p\bigl[e^{\beta G}\bigr].
\end{equation}
Here $u$ is the uniform law on $\mathcal S$, and $\mathcal L_0$ is taken after subtracting the constant $\hat\beta^{-1}\log|\mathcal S|$ from \eqref{eq:objective}. The first is the maximum-likelihood objective, $\beta^{-1}\mathrm{KL}(p\,\|\,q_G)$ up to a constant with $q_G\propto e^{-\beta G}$: its physical walkers are unbiased. The second is the fully flattened level, whose auxiliary target is $P_{G,0}=u$. These definitions agree with the corresponding interior limits, renormalized at zero, under the additional moment conditions in the proof of \cref{prop:kl-control}.

\begin{propbox}
\begingroup
\setlength{\abovedisplayskip}{3pt}
\setlength{\belowdisplayskip}{3pt}
\setlength{\abovedisplayshortskip}{0pt}
\setlength{\belowdisplayshortskip}{2pt}
\begin{proposition}[VESPER objective and minimizer]
\label[proposition]{thm:ladder}
Let $c_k>0$ with $\sum_{k=1}^Kc_k=1$, and define
\begin{equation}
\label{eq:physical-ladder}
\mathcal L_{\mathrm{VESPER}}[G]
=
\sum_{k=1}^K c_k\,\mathcal L_{\hat\beta_k}[G],
\end{equation}
with components given by \eqref{eq:objective}, or by \eqref{eq:VESPER_objective:lim} if $\hat\beta_1=0$ or $\hat\beta_K=\beta$. Under the integrability assumptions of \cref{app:proof-ladder}, $\mathcal L_{\mathrm{VESPER}}$ is convex in $G$, invariant under addition of a constant, and has minimizer set $\{G_\star+C:C\in\mathbb R\}$. In a realizable model family, every minimizer therefore represents $G_\star$ up to a constant. For a differentiable model $G_\theta$,
\begin{equation}
\label{eq:ladder-gradient}
\nabla_\theta\mathcal L_{\mathrm{VESPER}}[G_\theta]
=
\sum_{k=1}^K c_k
\left\{
\mathbb E_{\nu_{G_\theta,k}}
\!\left[\nabla_\theta G_\theta(s(x))\right]
-
\mathbb E_{P_{G_\theta,k}}
\!\left[\nabla_\theta G_\theta(s)\right]
\right\}.
\end{equation}
\end{proposition}
\endgroup
\end{propbox}

Every level, including strongly flattened ones, thus trains the same model directly, without reweighting between levels. The weights $c_k$ and exponents $\alpha_k$ leave the target unchanged but set how errors are weighted: near the minimizer, $\mathcal L_{\mathrm{VESPER}}[G]-\mathcal L_{\mathrm{VESPER}}[G_\star]\approx\tfrac{\beta}{2}\sum_k c_k\,\mathrm{Var}_{\pi_{\alpha_k}}(G-G_\star)$.

\subsection{Coupled sampling and optimization}
\label{sec:algorithm}

We estimate the expectations in \eqref{eq:ladder-gradient} with two populations at each level, all sharing $G_\theta$ (\cref{fig:diagram}). At level $k$, $n_k$ physical walkers evolve under the biased potential $U(x)-(1-\alpha_k)G_\theta(s(x))$, and $N_k$ auxiliary walkers evolve directly in CV space under the tempered potential $\alpha_kG_\theta(s)$:
\begin{equation}
\label{eq:sgdb}
\begin{aligned}
m_s\ddot s_t^{i,k}+\gamma_s\dot s_t^{i,k}
&=
-\alpha_k\nabla_sG_{\theta_t}(s_t^{i,k})
+
\sqrt{2\gamma_s\beta^{-1}}\,\eta_t^{i,k},
\\
m_x\ddot x_t^{j,k}+\gamma_x\dot x_t^{j,k}
&=
-\nabla_xU(x_t^{j,k})
+
(1-\alpha_k)\nabla_xG_{\theta_t}(s(x_t^{j,k}))
+
\sqrt{2\gamma_x\beta^{-1}}\,\tilde\eta_t^{j,k},
\end{aligned}
\end{equation}
where $m_s,m_x$ are masses, $\gamma_s,\gamma_x$ friction coefficients, and $\eta_t^{i,k},\tilde\eta_t^{j,k}$ independent white noises. The bias force on the molecular system is $\nabla_xG_\theta(s(x))=J_s(x)^\top\nabla_sG_\theta(s(x))$, with $J_s=\partial s/\partial x\in\mathbb R^{D\times d}$. For fixed $\theta$, these dynamics have configurational invariant laws $P_{G_\theta,k}$ and $\nu_{G_\theta,k}$ (\cref{app:proof-dynamics}), so their walker averages estimate the two expectations in \eqref{eq:ladder-gradient}. The parameters follow the corresponding gradient flow with $\gamma_\theta^{-1}$ the learning rate,
\begin{equation}
\label{eq:sgda}
\gamma_\theta\dot\theta_t
=
\sum_{k=1}^K c_k
\left[
\frac{1}{N_k}\sum_{i=1}^{N_k}
\nabla_\theta G_{\theta_t}(s_t^{i,k})
-
\frac{1}{n_k}\sum_{j=1}^{n_k}
\nabla_\theta G_{\theta_t}(s(x_t^{j,k}))
\right].
\end{equation}

Three features of this construction matter in practice.
\begin{itemize}[leftmargin=*]
\item \textbf{Timescales.} Since $\theta$ moves, the walkers track moving targets, and the averages in \eqref{eq:sgda} approximate the gradient when $\theta$ evolves more slowly than the walkers relax. \Cref{app:convergence} makes this quantitative for the overdamped dynamics: objective error and sampling lag decay together when the learning rate is below a threshold set by the log-Sobolev constants of the sampling laws and a bound on the parameter gradients, so no adiabatic limit is needed.
\item \textbf{Cost.} Auxiliary walkers live in the $D$-dimensional CV space and need one network evaluation per step, so $N_k$ can be large ($10^4$--$10^5$ in the molecular runs) with cost dominated by the biased MD.
\item \textbf{Persistence.} Auxiliary walkers persist between parameter updates, as in persistent contrastive divergence \citep{Tieleman2008}, staying close to equilibrium in the slowly changing tempered model.
\end{itemize}

\Cref{app:convergence} derives a conditional convergence bound for the coupled overdamped population dynamics without replica exchange. For $0<\alpha_k<1$, define $V_t=F(\theta_t)-F_*+H_t$, where $F=\beta\mathcal L_{\mathrm{VESPER}}$ and $H_t$ is the weighted KL sampling lag defined in \eqref{eq:tracking-entropy}. Under assumptions (C1)--(C4), for $0<\eta<a/B^2$,
\begin{equation}
\label{eq:main-convergence-rate}
 V_t\le e^{-\kappa t}V_0,
 \qquad \kappa=2\min\{\eta\mu,a-\eta B^2\}.
\end{equation}
Thus objective error and sampling lag decay together. Under (C1)--(C3), the same learning-rate restriction gives the time-averaged stationarity bound
\begin{equation}
\label{eq:main-stationarity-bound}
 \frac1T\int_0^T\|\nabla F(\theta_t)\|^2\,\dd t
 \le\frac{V_0}{\eta T}.
\end{equation}

In the implementation, \eqref{eq:sgda} is discretized: sampling epochs alternate with optimizer steps, and batches may be reused across several steps with within-batch importance weights (\cref{alg:vesper}, \cref{app:physics-trust}). When equilibrium configurations from $\varrho$ are available, the physical average can instead be computed from them with self-normalized weights (\cref{sec:data-learning}).

\subsection{Replica exchange across tempering levels}
\label{sec:replica-exchange}

The common-minimizer property does not require the levels to communicate. Sampling, however, benefits from it: strongly flattened levels cross barriers readily, while weakly flattened ones, including the physical level, may remain trapped. We therefore couple adjacent levels by replica exchange \citep{sugitaReplicaExchange1999,affentrangerNovelHamiltonianReplica2006,earlParallelTempering2005}, performed between parameter updates with $G_\theta$ held fixed. For a symmetric swap proposal between levels $i$ and $j$, the Metropolis acceptance probability for physical walkers is
\begin{equation}
\label{eq:main-exchange}
A_x(x^i,x^j)
=
\min\!\left\{1,\exp\!\left[(\hat\beta_i-\hat\beta_j)\bigl(G_\theta(s(x^i))-G_\theta(s(x^j))\bigr)\right]\right\},
\end{equation}
and $A_s(s^i,s^j)$ is identical with $s(x^{i})$, $s(x^{j})$ replaced by $s^i$, $s^j$.
Each swap preserves the fixed-model target product law $\prod_k\nu_{G_\theta,k}$ or $\prod_kP_{G_\theta,k}$, and hence the population-gradient identity (\cref{app:method-exchange}).
All physical replicas share the same potential $U$ and thermostat temperature, so the microscopic energy and Maxwell factors cancel from the exchange ratio. Both acceptance ratios therefore depend only on the learned free-energy differences in CV space. Given the CV values, their evaluation requires only two model values and no additional microscopic-energy evaluation. The tempering-level spacing is controlled by fluctuations of $G_\theta$, which can depend on the system and choice of CVs. Physical and auxiliary swaps are proposed independently.

\subsection{Related work}
\label{sec:related}

\paragraph{Adaptive-bias enhanced sampling.}
Umbrella sampling, adaptive biasing force, metadynamics and its well-tempered variant, and on-the-fly probability enhanced sampling accelerate MD with restraining or adaptively built CV-dependent biases during the simulation \citep{torrieNonphysicalSamplingDistributions1977a,darveCalculatingFreeEnergies2001a,darveAdaptiveBiasingForce2008,laioEscapingFreeenergyMinima2002a,barducciWellTemperedMetadynamicsSmoothly2008a,invernizziRethinkingMetadynamics2020}. Neural representations of the free energy or bias have been trained from mean forces, histograms, or both \citep{schneider2017nnfes,learning_FE_with_ANNs,guo2018funn,sevgen2020cff,zubietarico2025sobolev}. Reinforced dynamics couples such training to an uncertainty-gated bias \citep{zhang2018reinforced}. VESPER couples biased MD and persistent auxiliary CV walkers through one physical free-energy model, with all tempering levels contributing to an objective with a common minimizer (\cref{thm:ladder}). Its closest variational connection is to variationally enhanced sampling (VES) \citep{Valsson_2014}, in its well-tempered \citep{valsson2015welltempered} and neural \citep{bonatiNeuralNetworksbasedVariationally2019a} forms. With bias $V_\theta=-(1-\alpha)G_\theta$ and target $t_\theta\propto e^{-\alpha\beta G_\theta}$ held fixed during differentiation, the VES gradient equals $(1-\alpha)\nabla_\theta\mathcal L_{\hat\beta}[G_\theta]$ (\cref{app:ves-correspondence}). VESPER gives this instantaneous update an explicit objective in the physical free energy, with the current model directly determining both the molecular bias and the auxiliary target, without a separate recursive free-energy estimate. Neural VES also evaluates target averages by numerical integration or Monte Carlo. Here, persistent auxiliary dynamics form part of the coupled learning system (\cref{app:convergence}).

\paragraph{Replica ladders with CV biases.}
Replica exchange improves exploration by swapping configurations between replicas with different temperatures or Hamiltonians \citep{sugitaReplicaExchange1999,affentrangerNovelHamiltonianReplica2006,earlParallelTempering2005}. Bias-exchange metadynamics assigns different CVs to different replicas \citep{piana2007bemd}. Replica exchange with collective-variable tempering uses a ladder of well-tempered biases of increasing strength, which is the closest analogue of VESPER's physical population \citep{gilley2015rect}. VESPER uses the ladder for joint free-energy learning as well as configuration exchange: every level supplies its native gradient to one shared model, and the common-minimizer property allows these contributions to be combined without reweighting between levels (\cref{thm:ladder}).

\paragraph{Extended-variable methods.}
Temperature-accelerated MD, driven adiabatic free-energy dynamics, and related schemes couple the CVs to hot auxiliary variables through a stiff spring. They recover the free energy in the limit of adiabatic separation between auxiliary and molecular motion \citep{maraglianoTemperatureAcceleratedMethod2006a,abramsEfficientDirectGeneration2008a,rossoUseAdiabaticMolecular2002a,chenHeatingFloodingUnified2012a,awasthi2017tass}, possibly followed by mean-force reconstruction \citep{maraglianoSinglesweepMethodsFree2008a}. VESPER identifies the physical free energy through its variational minimizer, without requiring an adiabatic limit between auxiliary and molecular coordinates. Its auxiliary walkers explore the tempered model and communicate with the molecular system through shared parameters, without a spring coupling between the two populations.

\paragraph{Energy-based models and tempered objectives.}
VESPER connects tempered energy-based learning to molecular simulation through the prescribed CV map $s(x)$. Biased microscopic dynamics generate the tilted CV law, while persistent auxiliary walkers sample the tempered model directly in CV space \citep{Hinton2002,Tieleman2008}. Both populations train the same physical free-energy model without a pre-existing equilibrium dataset (\cref{app:proof-dynamics}). With $g=\beta G$ and $\gamma=\alpha-1$, the single-level objective takes the form of the pseudo-spherical contrastive divergence (PS-CD) objective \citep{yu2021pseudospherical}, built on the $\gamma$-cross-entropy \citep{fujisawa2008robust}, continued to negative $\gamma$, and admits a R\'enyi variational representation \citep{birrell2021renyi}. At $\alpha=1$ it reduces to maximum likelihood. For $0<\alpha<1$, VESPER uses the negative branch $\gamma=\alpha-1\in(-1,0)$. The factor $q_\theta^\gamma$ relatively upweights observations in regions of low model density and high model free energy. At the physical solution, both populations have CV densities proportional to $p^\alpha$, scaling CV free-energy barriers by $\alpha$ and increasing the relative weight of rare regions (\cref{app:related}). PS-CD estimates its positive term from a fixed dataset and its tempered model term by importance weighting model samples \citep{yu2021pseudospherical}. VESPER realizes both tilted laws through coupled dynamics, extending this variational principle to simultaneous molecular sampling and free-energy reconstruction (\cref{app:related}). Parallel tempering has also been used to train restricted Boltzmann machines \citep{desjardins2010tempered,cho2010parallel}, with only the untempered chain entering the gradient. In VESPER every level contributes.

\paragraph{Learned samplers.}
Boltzmann generators and non-equilibrium transport samplers learn maps or flows that sample a known energy \citep{noeBoltzmannGeneratorsSampling2019,dibakTemperatureSteerableFlows2022,invernizzi2023skipping,albergoNETS2025}, and annealed importance sampling and Jarzynski-type estimators weight paths between distributions \citep{nealAnnealedImportanceSampling2001,jarzynskiNonequilibriumEqualityFree1997,carbone2023jarzynski}. These methods target the full configurational distribution. VESPER instead targets a CV free energy through directly simulated tilted populations.

\section{Numerical results}\label{sec:experiments}

We evaluate VESPER on analytic landscapes and molecular systems to test free-energy reconstruction accuracy, convergence as the number of CVs increases, and exploration of conformational and chemical changes in complex MD systems. The analytic benchmarks provide exact references and controlled comparisons of tempering and replica exchange, while the molecular applications test the method in realistic MD. Complete simulation settings and numerical comparisons are given in the appendices. Molecular simulation times are per replica. Aggregate MD and population budgets are specified in \cref{tab:compute,app:fourwell-scaling}.

\subsection{Analytic landscapes: tempering and increasing dimension}\label{sec:toy-results}

\begin{figure}[t]
\centering
\includegraphics[width=\linewidth]{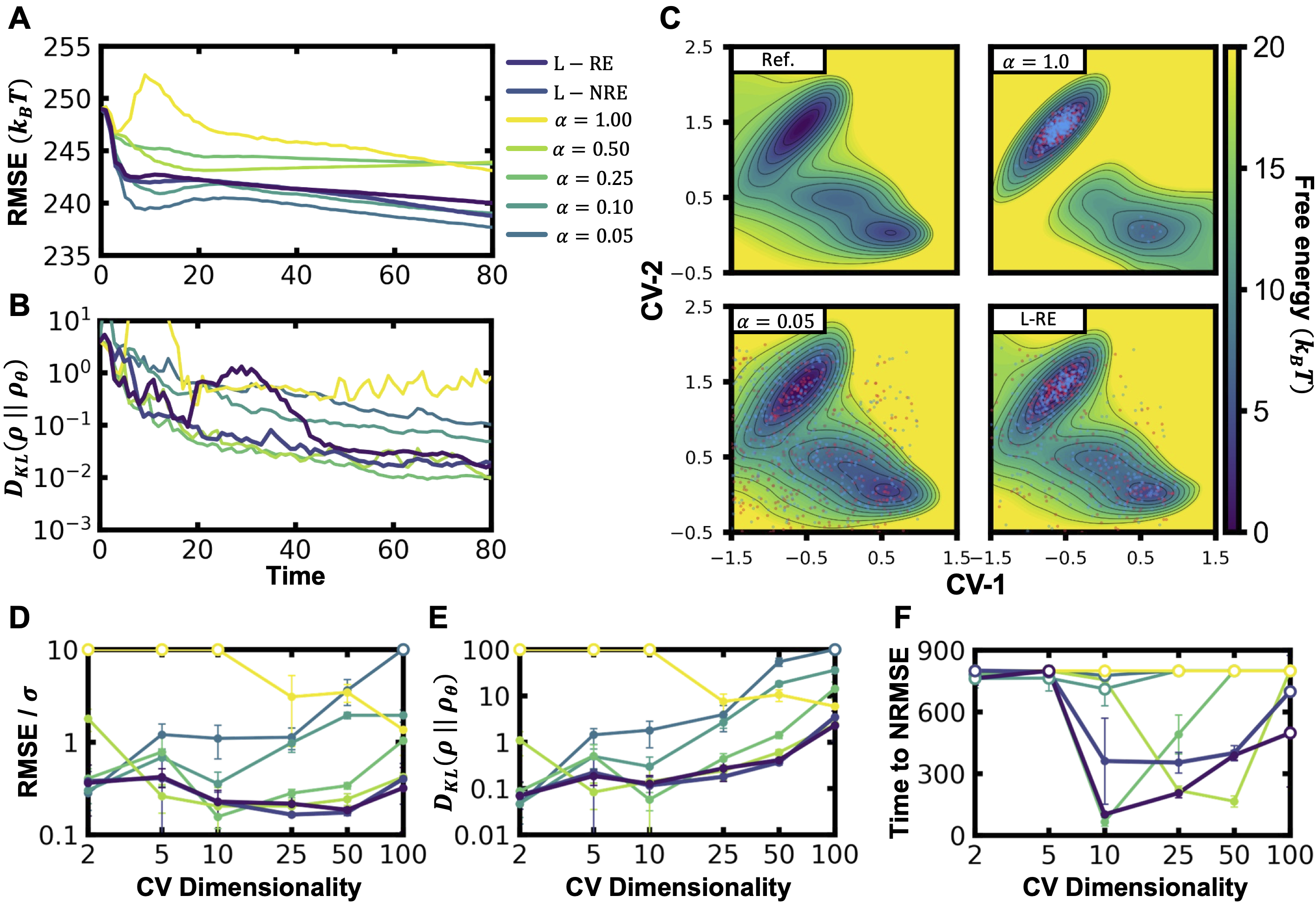}%
\caption{Analytic landscapes. (A) Full-evaluation-domain RMSE in $k_BT$ and (B) dimensionless $D_{\mathrm{KL}}(\rho\,\|\,\rho_\theta)$ against reduced time on M\"uller--Brown. Curves show means for single levels and a five-level ladder with (L-RE) and without (L-NRE) exchange, with $1{,}000$ walkers per population. (C) Reference and learned surfaces at $t=80$ on a common $0$--$20\,k_BT$ scale. Red and blue dots are physical and auxiliary walkers. (D--F) Four-well landscape in $d=2$--$100$ coordinates, with fixed hidden width and primary walker budget: (D) final SWA full-coordinate NRMSE, (E) $D_{\mathrm{KL}}(\rho\,\|\,\rho_\theta)$, and (F) reduced time to sustained EMA NRMSE below $0.25$. Points and error bars show means and sample SDs. Open circles in (D,E) mark values above the plotted range or non-finite divergences. Open markers in (F) indicate at least one run without a sustained crossing.}
\label{fig:res-toy}
\end{figure}

The M\"uller--Brown benchmark \citep{muellerLocationSaddle1979} tests how tempering balances accuracy in low-energy basins with reconstruction over a broader region of the landscape, including high-free-energy regions of low probability mass (\cref{fig:res-toy}A--C). Moderate tempering gives the lowest basin-sensitive KL divergence, while stronger tempering gives the lowest full-domain RMSE. The fixed exchange ladder balances these objectives, reaching a KL divergence of $0.017\pm0.013$ while keeping its mean RMSE within $1\%$ of the lowest tested value. Here the RMSE includes extrapolation beyond the simulation box into a high-energy rim, so it measures broad-domain reconstruction rather than barrier accuracy alone (\cref{app:mb-ablation}).

We then increase the dimension of a rotated four-well landscape from two to one hundred, keeping the hidden width, walker budget and update schedule fixed (\cref{fig:res-toy}D--F). The same tempering ladder remains competitive across the tested dimensions without selecting a different tempering strength for each problem. At one hundred dimensions, the exchange ladder gives the lowest mean normalized RMSE among the primary comparisons: $0.32\pm0.21$, compared with $0.42\pm0.05$ for the best tested single level. These results support using a shared ladder across dimensions, with its strongest advantages appearing in the more demanding exploration problem. The landscape, averaging conventions and larger-population controls are specified in \cref{app:fourwell-scaling}.

\subsection{Alanine dipeptide: comparison with established methods}\label{sec:adp}

\begin{figure}[t]
\centering
\includegraphics[width=\linewidth]{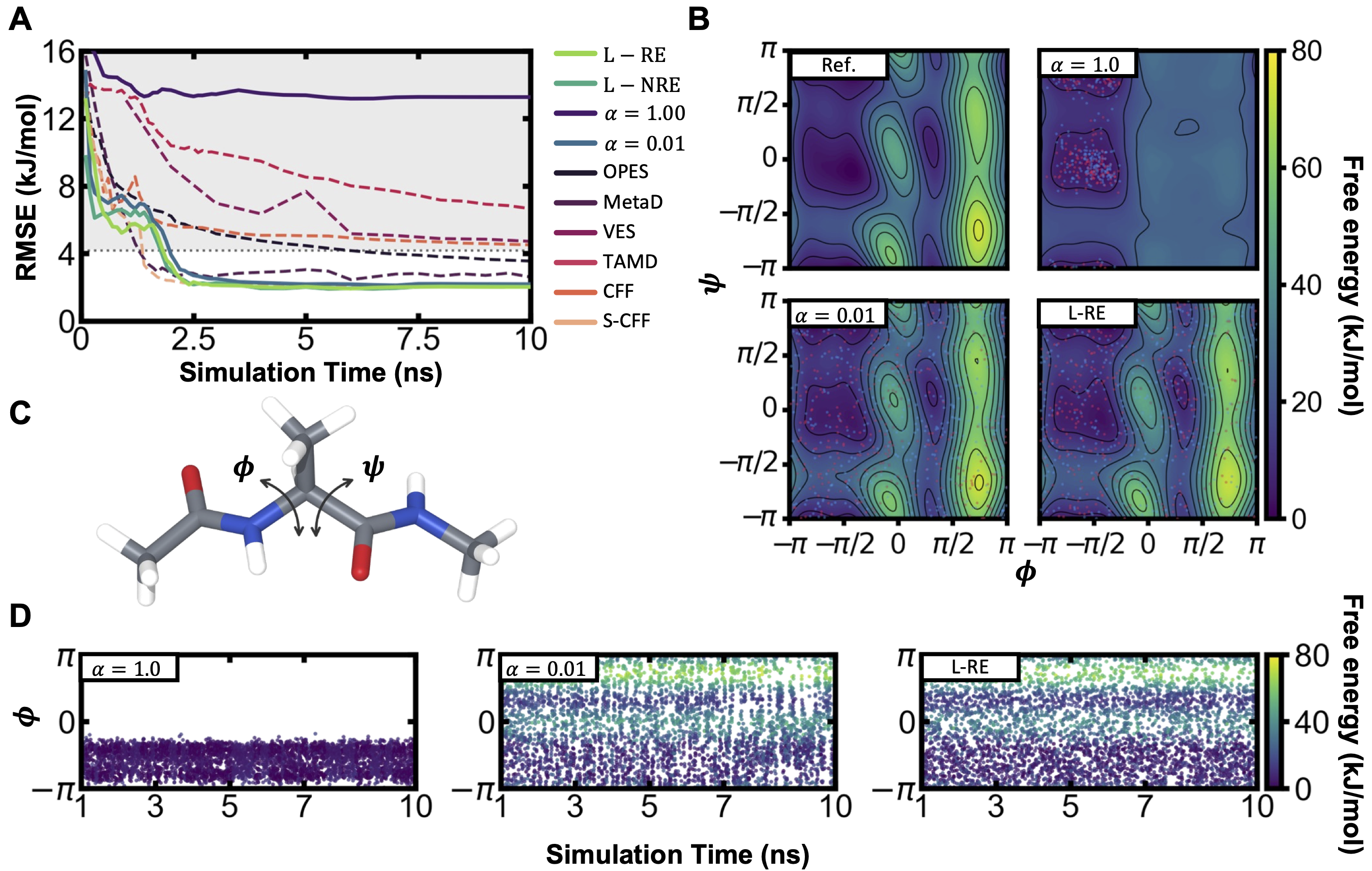}%
\caption{Alanine dipeptide. (A) RMSE against the reference surface \citep{zubietarico2025sobolev} over 0--10~ns per replica, averaged over three independent runs. VESPER curves are solid and baselines dashed. Gray shading marks errors above the chemical accuracy of 1~kcal\,mol$^{-1}$ (4.184~kJ\,mol$^{-1}$). L-RE, L-NRE: four-level ladders with and without exchange; OPES: on-the-fly probability enhanced sampling \citep{invernizziRethinkingMetadynamics2020}; MetaD: metadynamics \citep{laioEscapingFreeenergyMinima2002a}; VES: neural variationally enhanced sampling \citep{Valsson_2014,bonatiNeuralNetworksbasedVariationally2019a}; TAMD: temperature-accelerated MD \citep{maraglianoTemperatureAcceleratedMethod2006a,bajpaiInteroperableImplementationCollectivevariable2023}; CFF: combined force-frequency sampling \citep{sevgen2020cff}; S-CFF: its SIREN variant \citep{zubietarico2025sobolev}. (B) Reference (top left), and VESPER surfaces for $\alpha=1$ (top right), a single level at $\alpha=0.01$ (bottom left) and the exchange ladder (bottom right), on a common scale. Red and blue dots are physical and auxiliary walkers. (C) ADP with $\phi,\psi$ labeled \citep{humphrey1996vmd}. (D) $\phi$ of the physical walkers over 1--10~ns, colored by the learned free energy.}
\label{fig:res-adp}
\end{figure}

Alanine dipeptide (ADP) in explicit water tests whether VESPER can reconstruct a molecular free-energy surface with accuracy and convergence comparable to established enhanced-sampling methods. With the two backbone dihedrals $(\phi,\psi)$ as CVs, tempered VESPER explores the periodic landscape and recovers the reference surface \citep{zubietarico2025sobolev}, while the unbiased control remains concentrated in its initial basin (\cref{fig:res-adp}). At $10$~ns per replica, the exchange ladder reaches an RMSE of $2.02\pm0.24$~kJ\,mol$^{-1}$, comparable to $2.04\pm0.51$~kJ\,mol$^{-1}$ for SIREN-based combined force-frequency sampling (S-CFF), the strongest tested baseline at that horizon. A single tempered level also performs well. The ladders make only a small change to the endpoint error on this relatively simple low-dimensional landscape.

The VESPER mean error curves first fall below $1$~kcal\,mol$^{-1}$ and remain there through $10$~ns at approximately $2.0$~ns for a single level and $1.8$--$1.9$~ns for the ladders. Metadynamics and S-CFF reach this threshold earlier, at $1.3$--$1.4$~ns. VESPER therefore combines comparable late-time accuracy with rapid practical convergence, within about $2$~ns, on this benchmark. Full baseline comparisons and scoring conventions are given in \cref{app:physics-adp,app:comparison-plan,tab:reported}.

\subsection{Alanine tetrapeptide: consistency as the number of CVs increases}\label{sec:atp}

\begin{figure}[t]
\centering
\includegraphics[width=\linewidth]{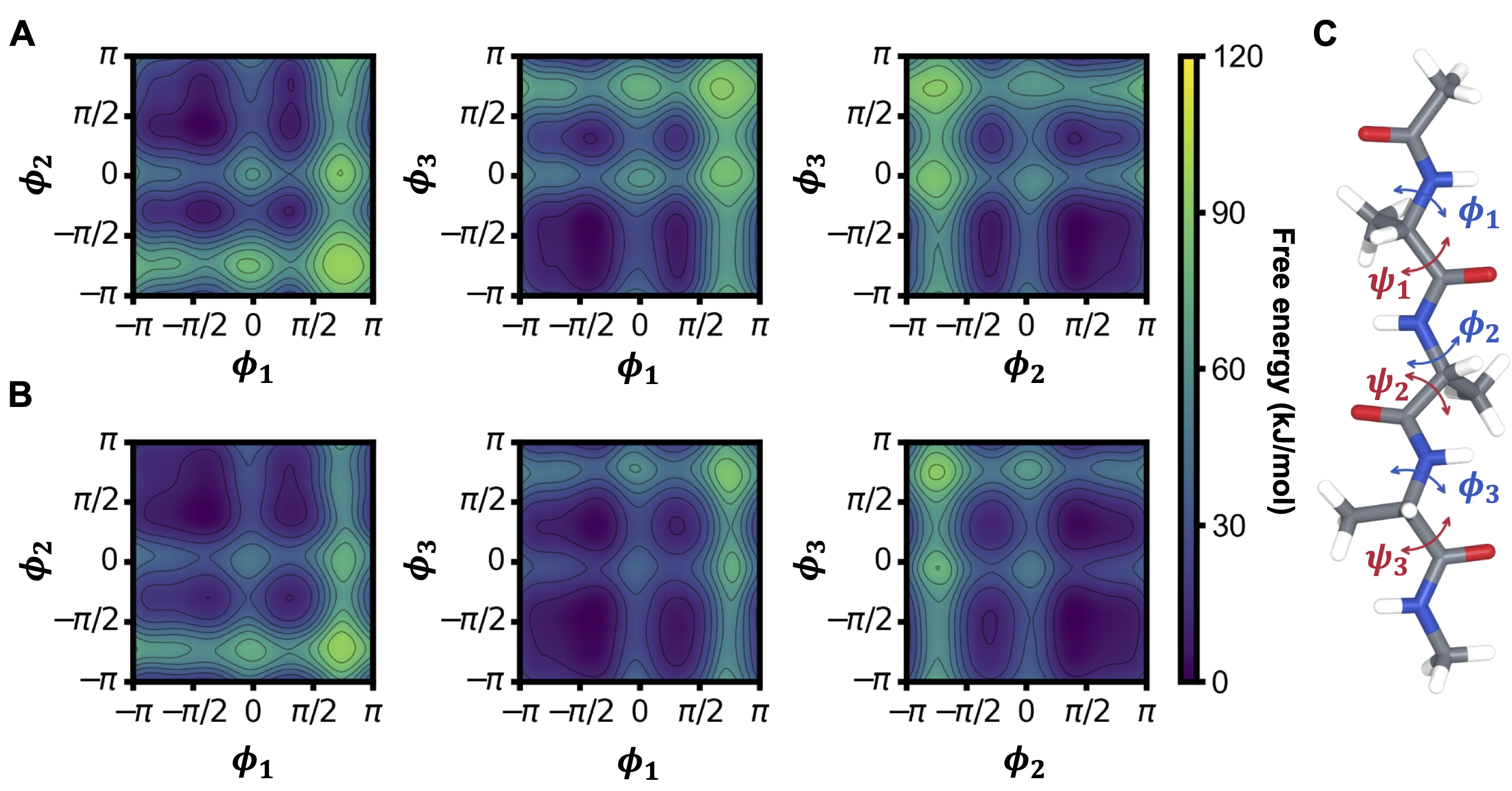}%
\caption{Alanine tetrapeptide: consistency across CV dimension. Two-dimensional Boltzmann marginals of (A) the three-CV and (B) the six-CV model, each from the mean surface of the four-level ladder with exchange. Columns: $(\phi_1,\phi_2)$, $(\phi_1,\phi_3)$, $(\phi_2,\phi_3)$. (C) ATP with $\phi$ bonds in blue and $\psi$ bonds in red.}
\label{fig:res-atp}
\end{figure}

Alanine tetrapeptide (ATP) tests whether VESPER retains a consistent molecular description when the number of biased backbone dihedrals increases from three to six. The two models differ only in their input layer, and their two-dimensional Boltzmann marginals retain similar broad structure (\cref{fig:res-atp}). Both six-CV ladder variants also produce more consistent learned shapes between $5$ and $10$~ns than the selected single-level model: the correlations with their own $10$~ns estimates are $0.913\pm0.012$ with exchange and $0.941\pm0.028$ without, compared with $0.804\pm0.179$ for the single level. This suggests that multiple learning levels can stabilize reconstruction over the observed interval in the higher-dimensional six-CV model. Without an independent converged six-CV reference, these comparisons establish dimensional and temporal consistency.

\subsection{Catalytic methane dehydrogenation on Ni(511)}\label{sec:reactive}

\begin{figure}[t]
\centering
\includegraphics[width=\linewidth]{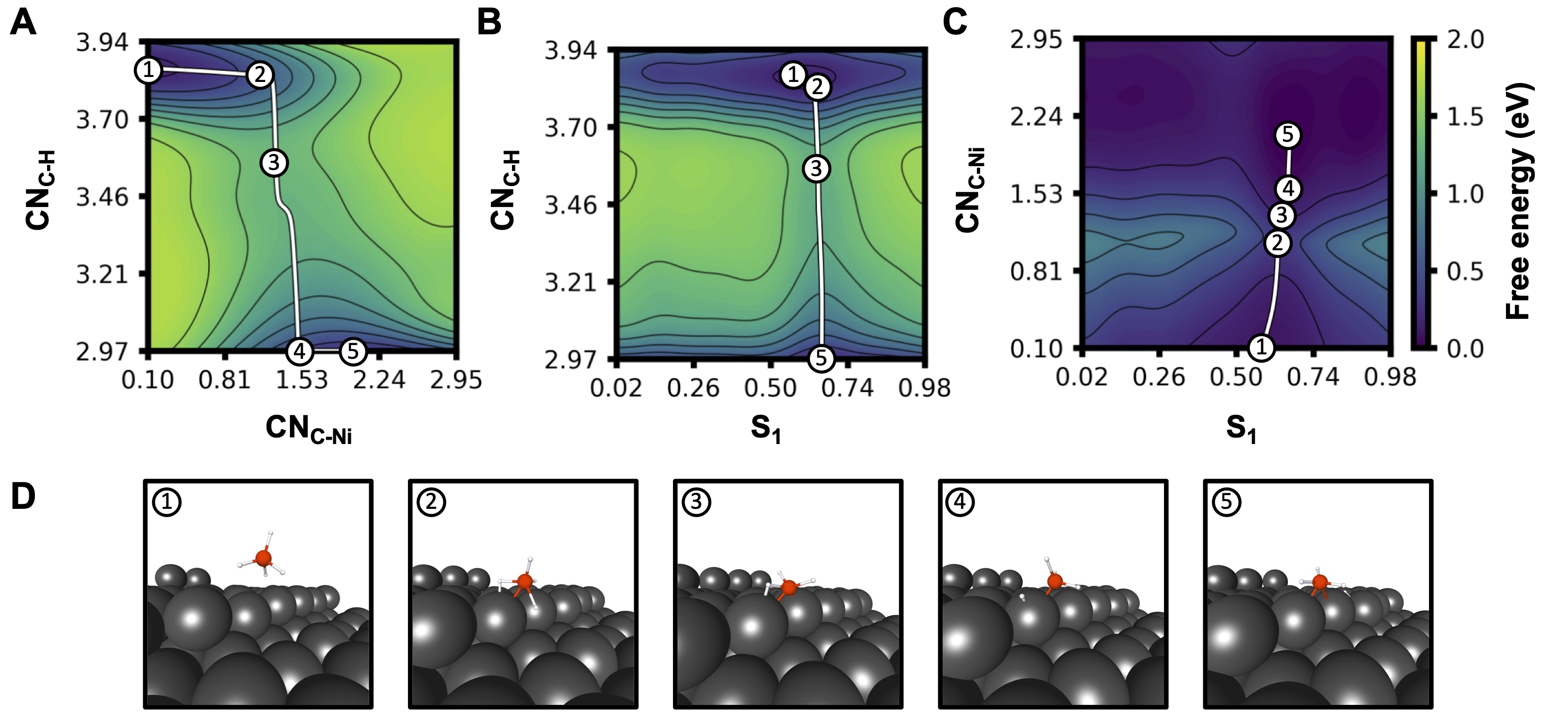}%
\caption{Methane on Ni(511): three-CV free energy. (A--C) Two-dimensional Boltzmann marginals in the planes (C--Ni, C--H coordination), ($S_1$, C--H coordination) and ($S_1$, C--Ni coordination). White line: bounded path on the mean surface obtained by a string-method calculation \citep{eSimplifiedImprovedString2007a} followed by constrained gradient-flow refinement, from reactant (1) through the saddle (3) to product (5). (D) Trajectory frames corresponding to the labeled points.}
\label{fig:res-cat}
\end{figure}

We next test VESPER on a reactive free-energy landscape described by a machine-learned interatomic potential. For methane on Ni(511) at $700$~K, a single tempered level learns a joint surface over the C--Ni and C--H coordination numbers and the carbon atom's lateral position. Its two-CV Boltzmann marginals recover the physisorbed and dissociated basins and the transition region of the lower-dimensional reference \citep{jinFreeEnergyLandscapesSurface2025} (\cref{fig:res-cat}).

The differentiable neural representation also supports post hoc analysis of the joint three-CV landscape using its gradients. A string-method path calculation \citep{eSimplifiedImprovedString2007a}, followed by constrained gradient-flow refinement, connects reactant and product through a fitted saddle. The bounded reactant-to-saddle free-energy difference is $1.12\pm0.25$~eV, consistent with the reference. Including the lateral position directly in the model locates the fitted saddle near the step-edge atoms. Whereas \citet{jinFreeEnergyLandscapesSurface2025} obtained an additional spatial free-energy description through post hoc Spectral-ABF analysis of existing trajectories, VESPER learns the joint three-CV surface during sampling. The route, averaging and projection conventions are specified in \cref{app:submission-applications}.

\subsection{Chignolin: exploration with seventeen CVs}\label{sec:chignolin}

\begin{figure}[t]
\centering
\includegraphics[width=\linewidth]{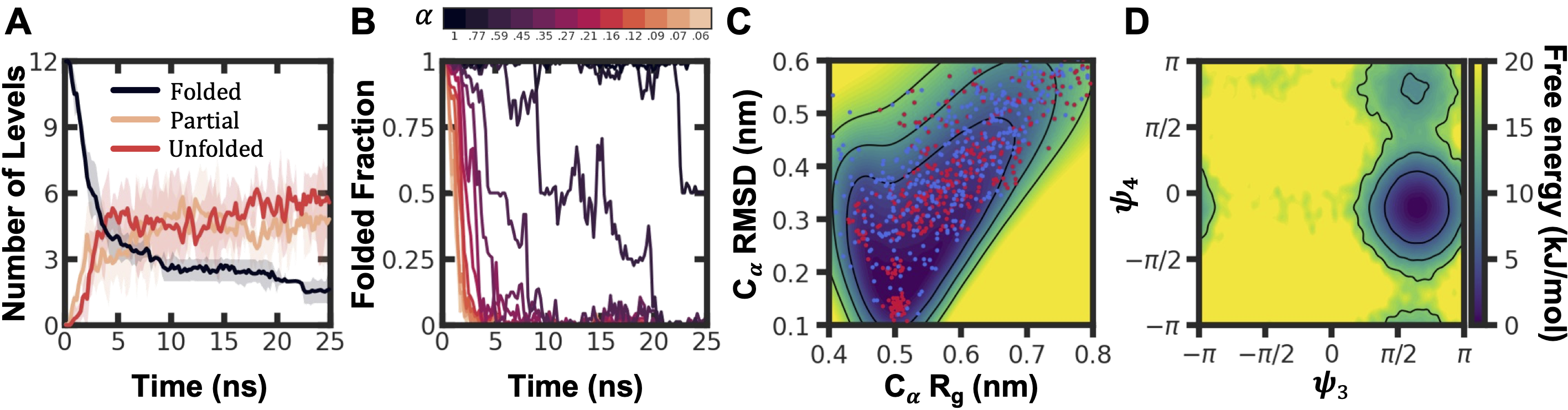}%
\caption{Chignolin with seventeen dihedral CVs (four independent runs, twelve levels from $\alpha=1$ to $1/18$). (A) Mean number of levels holding folded (C$_\alpha$ RMSD below $0.20$~nm), intermediate (labeled Partial) or unfolded (above $0.40$~nm) configurations. Shading shows the minimum-to-maximum range across runs. (B) Mean folded fraction at each tempering level. In A and B, occupancies are averaged in $0.2$~ns bins and across the four runs. Both panels display the first $25$~ns per replica. (C) Free-energy projection at the end of the run from unbiased-level auxiliary walkers reconstructed on a fixed backbone, in C$\alpha$ radius of gyration and RMSD (dots: MD frames, red, and auxiliary walkers, blue, from all levels). (D) Dihedral marginal in $(\psi_3,\psi_4)$ from the same unbiased-level auxiliary walkers. Contours every $5$~kJ\,mol$^{-1}$.}
\label{fig:res-chig}
\end{figure}

Finally, we test whether one shared neural free energy can support exploration in the high-dimensional space of chignolin's seventeen backbone dihedrals. For CLN025 in explicit water, a twelve-level ladder continuously trains the joint model during replica exchange, with every level contributing to learning (\cref{fig:res-chig}). The representation is not restricted to a sum of pairwise terms, and the target is not prescribed as a folded/unfolded mixture, extending earlier work on this system \citep{shafferEnhancedTargeted2016}.

Starting from folded configurations, VESPER explores RMSD-defined folded and unfolded regions. Exploration broadens along the ladder while the unbiased level remains predominantly folded. Auxiliary-sample projections place the free-energy minimum in the folded region and represent a broader range of conformations also visited by MD. VESPER thus learns a joint free energy while exploring conformational changes with seventeen CVs.

\section{Discussion and conclusion}\label{sec:conclusion}

We introduced VESPER, which casts free-energy learning and enhanced sampling as a single optimization problem. The free energy is parameterized by a neural network $G_\theta$ that biases the MD, drives tempered auxiliary walkers in CV space, and is trained on the difference between its mean parameter gradients over the auxiliary walkers and the biased MD configurations. This difference estimates the gradient of one variational objective, so the simulation generates its own training data as the model improves. Working in an extended space of molecular configurations and auxiliary walkers, VESPER minimizes an explicit objective whose minimizer is the physical free energy itself. The free energy is therefore learned directly, rather than recovered from a bias optimized toward a fixed or recursively updated target, and no adiabatic separation between auxiliary and molecular coordinates is required. Because the objective is convex and every tempering level shares the same global minimizer, all levels train one model without reweighting between them. Strongly flattened levels that cross barriers readily therefore work alongside weakly tempered levels, and when the ladder includes the physical level, $\alpha=1$, the ladder objective bounds the KL divergence of the model from the true CV distribution. Replica exchange moves configurations along the ladder using only learned free-energy differences in CV space, with no additional evaluations of the microscopic potential. The coupled sampling and learning dynamics also carry a conditional convergence guarantee.

VESPER is effective across a wide range of problems. The analytic landscapes provide exact references for controlled tests. On the M\"uller--Brown potential, the exchange ladder balances accuracy in the low-energy basins with reconstruction of the broader landscape. On a four-well landscape whose dimension grows from two to one hundred, one fixed ladder stays competitive without retuning and shows its clearest advantage at one hundred dimensions. Alanine dipeptide is a standard molecular benchmark with a reference surface for direct comparison. Here VESPER matches or improves on the accuracy of established enhanced-sampling methods, reaching the end-point error of the best tested baseline, and converges within $2$~ns. Alanine tetrapeptide tests whether the learned description stays consistent as more CVs are biased, and the Boltzmann marginals learned with three and six CVs agree. Methane dehydrogenation on Ni(511) extends the tests from conformational to chemical change, using a machine-learned interatomic potential. VESPER learns a reactive three-CV landscape whose two-CV marginals recover the basins and transition region of the reference. Because the model is a differentiable neural network, a string-method calculation with gradient-flow refinement then maps the reaction path directly on the joint three-CV surface learned during sampling. Chignolin tests exploration in a realistic high-dimensional CV space. While biasing seventeen backbone dihedrals, one shared model drives repeated unfolding and refolding. These results connect online free-energy learning to enhanced sampling across conformational and chemical changes. Molecular comparisons retain their per-replica cost: a $K$-level ladder uses $K$ times the aggregate MD of a single replica at the same horizon.

VESPER directly accelerates barrier crossing along the chosen CVs. Orthogonal slow motions can still limit exploration by the physical walkers. Our convergence analysis assumes overdamped dynamics and a Polyak--\L{}ojasiewicz condition and excludes exchange, batch reuse and practical optimizers. The high-dimensional molecular results establish self-consistency and exploration without converged free-energy references. The stability controls for batch reuse (\cref{app:physics-trust}) add per-system hyperparameters. Natural extensions are joint CV learning, adaptive level weights, and combining online sampling with existing trajectories (\cref{sec:data-learning}).

\subsubsection*{Reproducibility statement}
Proofs, protocols, scoring conventions and the algorithm are given in the appendices. VESPER is integrated into PySAGES \citep{zubietaricoPySAGESFlexibleAdvanced2024}, available at {\hypersetup{urlcolor=blue}\url{https://github.com/SSAGESLabs/PySAGES}}. All code, run configurations and analysis scripts will be released publicly at {\hypersetup{urlcolor=blue}\url{https://github.com/JayLKap/vesper}}.
Generative AI tools assisted with code implementation, experiment execution, figure preparation,
literature review, and manuscript editing under author supervision. The authors reviewed the
AI-assisted work and take full responsibility for the final content.

\bibliography{vesper_arxiv}
\bibliographystyle{arxiv}

\clearpage
\appendix
\section{Proofs}\label{app:method-details}\label{app:proofs}

We retain the notation of the main text, with $G$ in energy units and $\alpha=\hat\beta/\beta$. When considering a single tempering level, we write $P_G$, $\nu_G$, and $Q_G$ for $P_{G,\hat\beta}$, $\nu_{G,\hat\beta}$, and $Q_{G,\hat\beta}$, respectively.

\theoremstyle{plain}
\newtheorem{siproposition}{Proposition}
\renewcommand{\thesiproposition}{S\arabic{siproposition}}
\newtheorem{sitheorem}[siproposition]{Theorem}
\newtheorem{sicorollary}[siproposition]{Corollary}
\crefname{siproposition}{proposition}{propositions}
\Crefname{siproposition}{Proposition}{Propositions}
\crefname{sitheorem}{theorem}{theorems}
\Crefname{sitheorem}{Theorem}{Theorems}
\crefname{sicorollary}{corollary}{corollaries}
\Crefname{sicorollary}{Corollary}{Corollaries}

\subsection{Convexity and the physical free-energy minimizer}\label{app:proof-convexity}

Let $p$ be the physical CV density in \eqref{eq:physical-free-energy}. We work on its support $\mathcal S$, where $p>0$ almost everywhere, and for the chosen $0<\alpha<1$ assume that
\begin{equation}\label{eq:si-integrability}
 C_\alpha=\int_{\mathcal S}p(s)^\alpha\,\dd s<\infty.
\end{equation}
We consider finite-valued measurable free energies $G$ for which the two normalizers
\begin{equation}\label{eq:si-normalizers}
 A_G=\int_{\mathcal S}e^{-\hat\beta G(s)}\,\dd s,
 \qquad B_G=\int_{\mathcal S}p(s)e^{(\beta-\hat\beta)G(s)}\,\dd s
\end{equation}
are finite and positive.

The auxiliary density, biased microscopic density and its CV marginal are
\begin{equation}\label{eq:physical-pair-supp}
 P_G(s)=\frac{e^{-\hat\beta G(s)}}{A_G},\qquad
 \nu_G(x)=\frac{e^{(\beta-\hat\beta)G(s(x))}\varrho(x)}{B_G},
\end{equation}
\begin{equation}\label{eq:physical-cv-law}
 Q_G(s)=\frac{p(s)e^{(\beta-\hat\beta)G(s)}}{B_G}.
\end{equation}

\begin{siproposition}[Convexity and the free-energy minimizer]\label{thm:consistency}
The functional in \eqref{eq:objective} is convex in $G$, invariant under addition of a constant, and has minimizer set $\{G_\star+C\}$. Its minimum is
\begin{equation}\label{eq:si-minimum-value}
 \mathcal L_{\hat\beta}[G_\star]
 =\frac{\log C_\alpha}{\beta\alpha(1-\alpha)}.
\end{equation}
At this minimum, $P_{G_\star}=Q_{G_\star}=p^\alpha/C_\alpha$.
\end{siproposition}

\begin{proof}
Each term of \eqref{eq:objective} is convex by H\"older's inequality. Adding a constant $c$ to $G$ changes the two terms by $-c$ and $c$, so their sum is unchanged. To identify the minimizer, differentiate in a bounded direction $h$:
\begin{align}
 D\mathcal L_{\hat\beta}[G](h)
 &=\mathbb E_{Q_G}h-\mathbb E_{P_G}h,
 \label{eq:si-first-variation}\\
 D^2\mathcal L_{\hat\beta}[G](h,h)
 &=\hat\beta\operatorname{Var}_{P_G}(h)
  +(\beta-\hat\beta)\operatorname{Var}_{Q_G}(h)\ge0.
 \label{eq:si-second-variation}
\end{align}
The first variation vanishes for every such $h$ exactly when $P_G=Q_G$. Using \eqref{eq:physical-pair-supp} and \eqref{eq:physical-cv-law}, this condition becomes
\[
 \frac{e^{-\hat\beta G(s)}}{A_G}
 =\frac{p(s)e^{(\beta-\hat\beta)G(s)}}{B_G}
 \quad\Longleftrightarrow\quad
 G(s)=-\beta^{-1}\log p(s)+C.
\]

To see that this candidate attains the minimum, apply H\"older's inequality once more:
\[
 C_\alpha
 =\int_{\mathcal S}
 (e^{-\hat\beta G})^{1-\alpha}
 (p e^{(\beta-\hat\beta)G})^\alpha\,\dd s
 \le A_G^{1-\alpha}B_G^\alpha.
\]
Taking logarithms gives the lower bound in \eqref{eq:si-minimum-value}. Equality holds precisely when the two normalized factors coincide, hence precisely at $G=G_\star+C$. This candidate is admissible by \eqref{eq:si-integrability}, and substitution gives $P_{G_\star}=Q_{G_\star}=p^\alpha/C_\alpha$. Finally, applying the chain rule to \eqref{eq:si-first-variation} gives the parameter gradient in \eqref{eq:physical-gradient}, provided differentiation under the integrals is justified.
\end{proof}

\subsection{KL control and endpoint limits}\label{app:proof-kl}

\begin{siproposition}[Loss and KL control]\label{prop:kl-control}
Let $\pi_\alpha=p^\alpha/C_\alpha$. Under the assumptions of \cref{thm:consistency} and integrability of $G$ and $\log p$ under $\pi_\alpha$,
\begin{equation}\label{eq:si-kl-decomposition}
 \mathcal L_{\hat\beta}[G]-\mathcal L_{\hat\beta}[G_\star]
 =\frac{\operatorname{KL}(\pi_\alpha\|P_G)}{\hat\beta}
  +\frac{\operatorname{KL}(\pi_\alpha\|Q_G)}{\beta-\hat\beta}.
\end{equation}
In particular,
\begin{equation}\label{eq:si-kl-control}
 \operatorname{KL}(\pi_\alpha\|P_G)
 \le\hat\beta\bigl(\mathcal L_{\hat\beta}[G]
                         -\mathcal L_{\hat\beta}[G_\star]\bigr),
 \qquad
 \operatorname{KL}(\pi_\alpha\|Q_G)
 \le(\beta-\hat\beta)\bigl(\mathcal L_{\hat\beta}[G]
                         -\mathcal L_{\hat\beta}[G_\star]\bigr).
\end{equation}
At $\hat\beta=\beta$, assume $Z_G=\int e^{-\beta G}\,\dd s<\infty$, $\mathbb E_p|G|<\infty$, and $\mathbb E_p|\log p|<\infty$. With $q_G=e^{-\beta G}/Z_G$, the extension in \eqref{eq:VESPER_objective:lim} satisfies
\begin{equation}\label{eq:si-endpoint-kl}
 \mathcal L_\beta[G]-\mathcal L_\beta[G_\star]
 =\beta^{-1}\operatorname{KL}(p\|q_G).
\end{equation}
At $\hat\beta=0$ on compact $\mathcal S$, assume $\mathbb E_u|G|<\infty$, $\mathbb E_u|\log p|<\infty$, and $0<B_{G,0}:=\mathbb E_pe^{\beta G}<\infty$. With $Q_{G,0}=p e^{\beta G}/B_{G,0}$, the zero-level objective satisfies
\begin{equation}\label{eq:si-zero-endpoint-kl}
 \mathcal L_0[G]-\mathcal L_0[G_\star]
 =\beta^{-1}\operatorname{KL}(u\|Q_{G,0}).
\end{equation}
\end{siproposition}

\begin{proof}
Substituting the expressions for $P_G$ and $Q_G$ into the two divergences gives
\begin{align*}
 \operatorname{KL}(\pi_\alpha\|P_G)
 &=\mathbb E_{\pi_\alpha}\log\pi_\alpha
   +\hat\beta\mathbb E_{\pi_\alpha}G+\log A_G,\\
 \operatorname{KL}(\pi_\alpha\|Q_G)
 &=\mathbb E_{\pi_\alpha}\log\pi_\alpha
   -\mathbb E_{\pi_\alpha}\log p
   -(\beta-\hat\beta)\mathbb E_{\pi_\alpha}G+\log B_G.
\end{align*}
Dividing by $\hat\beta$ and $\beta-\hat\beta$, respectively, and adding cancels the expectations of $G$. Since $\log\pi_\alpha=\alpha\log p-\log C_\alpha$, the remaining terms are
$\mathcal L_{\hat\beta}[G]-\log C_\alpha/[\beta\alpha(1-\alpha)]$.
This proves \eqref{eq:si-kl-decomposition}, and nonnegativity of either divergence gives \eqref{eq:si-kl-control}.

At the physical endpoint, the definition in \eqref{eq:VESPER_objective:lim} gives directly
\[
 \operatorname{KL}(p\|q_G)
 =\mathbb E_p\log p+\beta\mathbb E_pG+\log Z_G
 =\beta\mathcal L_\beta[G]+\mathbb E_p\log p.
\]
Since $q_{G_\star}=p$, subtracting the same identity at $G_\star$ proves \eqref{eq:si-endpoint-kl}.

To recover this endpoint as a limit of the interior objectives, assume in addition that $\int e^{-\hat\beta_0G}\,\dd s<\infty$ for some $0<\hat\beta_0<\beta$ and $\mathbb E_pe^{\varepsilon_0G}<\infty$ for some $\varepsilon_0>0$. The bound $e^{-\hat\beta G}\le e^{-\hat\beta_0G}+e^{-\beta G}$ then gives $\hat\beta^{-1}\log A_G\to\beta^{-1}\log Z_G$ by dominated convergence. Writing $\varepsilon=\beta-\hat\beta\downarrow0$, the difference quotient $(e^{\varepsilon G}-1)/\varepsilon$ decreases to $G$ and lies between the integrable functions $G$ and $(e^{\varepsilon_0G}-1)/\varepsilon_0$. Thus $\mathbb E_pe^{\varepsilon G}=1+\varepsilon\mathbb E_pG+o(\varepsilon)$, so the second term of \eqref{eq:objective}, $\varepsilon^{-1}\log B_G=\varepsilon^{-1}\log\mathbb E_pe^{\varepsilon G}$, tends to $\mathbb E_pG$ as required.

At the zero endpoint, direct expansion gives
\[
 \operatorname{KL}(u\|Q_{G,0})
 =\beta\mathcal L_0[G]-\mathbb E_u\log p-\log|\mathcal S|.
\]
Since $\beta\mathcal L_0[G_\star]=\mathbb E_u\log p+\log|\mathcal S|$, subtraction proves \eqref{eq:si-zero-endpoint-kl}. Equality holds exactly when $Q_{G,0}=u$, or $G=G_\star+C$ almost everywhere. For the renormalized limit $\hat\beta\downarrow0$, additionally assume $\int e^{-\hat\beta_0G}\,\dd s<\infty$ for some $0<\hat\beta_0<\beta$. The same difference-quotient argument under $u$ gives $\hat\beta^{-1}\log\mathbb E_u e^{-\hat\beta G}\to-\mathbb E_uG$, while dominated convergence gives $\mathbb E_pe^{(\beta-\hat\beta)G}\to B_{G,0}$. These are the two terms of the renormalized limit.
\end{proof}

\subsection{Dual formulation and coupled dynamics}\label{app:proof-dynamics}

For a probability density $\rho$ on $\mathcal S$, define its entropy relative to the CV reference measure by
$\operatorname{Ent}_{\dd s}(\rho)=\int_{\mathcal S}\rho\log\rho\,\dd s$.
Let $\nu$ be a probability law on configuration space. Define
\begin{equation}\label{eq:si-dual}
 \mathcal J(G,\rho,\nu)
 =\mathbb E_\nu[G\circ s]-\mathbb E_\rho G
  -\frac{\operatorname{Ent}_{\dd s}(\rho)}{\hat\beta}
  -\frac{\operatorname{KL}(\nu\|\varrho)}{\beta-\hat\beta}.
\end{equation}

\begin{siproposition}[Dual formulation and stationary sampling laws]\label{thm:pushforward}
For $0<\hat\beta<\beta$ and admissible $G$,
\begin{equation}\label{eq:saddle}
 \mathcal L_{\hat\beta}[G]=\sup_{\rho,\nu}\mathcal J(G,\rho,\nu),
 \qquad
 \min_G\mathcal L_{\hat\beta}[G]
 =\min_G\sup_{\rho,\nu}\mathcal J(G,\rho,\nu).
\end{equation}
The maximizing laws are $P_G$ and $\nu_G$, and the pushforward of $\nu_G$ is $Q_G$. Suppose the potentials and CV map are smooth, the Langevin equations are well posed, and the boundaries are periodic, reflecting, or confining with sufficient decay. The fixed-model dynamics in \eqref{eq:sgdb} then preserve the configurational laws $P_G$ and $\nu_G$.
\end{siproposition}

\begin{proof}
The dual identity follows by expanding the relative entropies:
\begin{equation}\label{eq:si-dual-gap}
 \mathcal J(G,\rho,\nu)
 =\mathcal L_{\hat\beta}[G]
  -\frac{\operatorname{KL}(\rho\|P_G)}{\hat\beta}
  -\frac{\operatorname{KL}(\nu\|\nu_G)}{\beta-\hat\beta}.
\end{equation}
The two KL terms are nonnegative and vanish at $\rho=P_G$ and $\nu=\nu_G$, proving \eqref{eq:saddle}. The expansion is immediate when all terms are finite, and the Gibbs variational formula gives the same supremum otherwise.

For fixed $\rho$ and $\nu$, differentiating $\mathcal J$ with respect to $\theta$, under the integrals, gives
$\mathbb E_\nu\nabla_\theta G_\theta\circ s-\mathbb E_\rho\nabla_\theta G_\theta$.
At the maximizing laws this is $\nabla_\theta\mathcal L_{\hat\beta}[G_\theta]$. Taking the negative derivative at each level, weighting by $c_k$, and replacing the expectations by walker averages gives the parameter update in \eqref{eq:sgda}. Meanwhile, Wasserstein ascent in $\rho$ and $\nu$, with mobilities $\alpha/\gamma_s$ and $(1-\alpha)/\gamma_x$, gives
\begin{align}
 \partial_t\rho
 &=\gamma_s^{-1}\nabla_s\cdot
       (\alpha\rho\nabla_sG+\beta^{-1}\nabla_s\rho),
 \label{eq:si-dual-rho-flow}\\
 \partial_t\nu
 &=\gamma_x^{-1}\nabla_x\cdot
       \bigl(\nu\nabla_x[U-(1-\alpha)G\circ s]
                          +\beta^{-1}\nabla_x\nu\bigr).
 \label{eq:si-dual-nu-flow}
\end{align}
Their particle realizations are the overdamped sampling equations. Adding momenta and Langevin thermostats yields the inertial realization in \eqref{eq:sgdb} with the same configurational targets.

The CV marginal of $\nu_G$ follows from the definition of $p$: for any bounded test function $h$,
\[
 \int_{\mathcal X}h(s(x))\nu_G(x)\,\dd x
 =\frac{1}{B_G}\int_{\mathcal S}h(s)p(s)e^{(\beta-\hat\beta)G(s)}\,\dd s.
\]
Thus $s_\#\nu_G=Q_G$. To check invariance, note that the potentials $\alpha G$ and $U-(1-\alpha)G\circ s$ at inverse temperature $\beta$ have Gibbs factors $e^{-\hat\beta G}$ and $e^{-\beta U+(\beta-\hat\beta)G\circ s}$. These make the probability currents in \eqref{eq:si-dual-rho-flow} and \eqref{eq:si-dual-nu-flow} vanish. For inertial dynamics, the corresponding phase-space laws include Maxwell momentum factors: Hamiltonian transport preserves energy and phase volume, while friction and noise preserve the Maxwell factors. Under ergodicity, time averages therefore recover the required expectations.

The sampling laws also hold at $\hat\beta=\beta$, although the dual functional in \eqref{eq:si-dual} is singular there. The auxiliary potential is then $G$, with invariant density $P_{G,\beta}\propto e^{-\beta G}$, and the unbiased physical dynamics preserve $\varrho$.
\end{proof}

\subsection{The common-minimizer ladder}\label{app:proof-ladder}

\begin{siproposition}[Weighted excess loss and level gradients]\label{prop:ladder-si}
Assume \cref{thm:consistency} at every interior level and the moment and normalizer conditions of \cref{prop:kl-control} at an endpoint level. For $c_k>0$ with $\sum_kc_k=1$, the ladder functional is convex and its minimizer set is $\{G_\star+C\}$. Its parameter gradient is given by \eqref{eq:ladder-gradient}. These statements imply \cref{thm:ladder}.
\end{siproposition}

\begin{proof}
For each level, \cref{thm:consistency,prop:kl-control} give
\[
 \Delta_k[G]:=\mathcal L_{\hat\beta_k}[G]
              -\mathcal L_{\hat\beta_k}[G_\star]\ge0,
 \qquad \Delta_k[G]=0\ \Longleftrightarrow\ G=G_\star+C.
\]
This includes both endpoints by \eqref{eq:si-endpoint-kl} and \eqref{eq:si-zero-endpoint-kl}. The excess of the combined objective is $\sum_kc_k\Delta_k[G]$. Since every $c_k$ is positive, this sum vanishes exactly when every component vanishes, proving the common-minimizer claim. Each component is convex and invariant under constant shifts, so the weighted sum has both properties as well. At either endpoint, these properties follow directly from \eqref{eq:VESPER_objective:lim}: the log-partition term is convex, the expectation is linear, and their shifts cancel. If the model family contains $G_\star+C$, it attains the same minimum.

Finally, differentiating the finite sum gives
\[
 \nabla_\theta\mathcal L_{\mathrm{VESPER}}[G_\theta]
 =\sum_kc_k\nabla_\theta\mathcal L_{\hat\beta_k}[G_\theta].
\]
Substitution of \eqref{eq:physical-gradient} proves \eqref{eq:ladder-gradient}. At the physical endpoint, differentiating \eqref{eq:VESPER_objective:lim} gives $\mathbb E_p\nabla_\theta G_\theta-\mathbb E_{P_{G_\theta,\beta}}\nabla_\theta G_\theta$, assuming differentiation under the integral is justified. This has the same form, since $\nu_{G,K}=\varrho$ and $s_\#\varrho=p$. At the zero endpoint the gradient is $\mathbb E_{Q_{G_\theta,0}}\nabla_\theta G_\theta-\mathbb E_u\nabla_\theta G_\theta$, again assuming justified differentiation, with $P_{G_\theta,0}=u$ and $s_\#\nu_{G_\theta,0}=Q_{G_\theta,0}$. Thus each level contributes expectations under its own sampling laws, with coefficient $c_k$.
\end{proof}

\subsection{Replica-exchange acceptance}\label{app:method-exchange}

\begin{siproposition}[Exchange preserves the level product laws]\label{prop:exchange}
Fix $G$ for which all level distributions are normalizable, and consider replicas of the same physical system at inverse temperature $\beta$, with auxiliary inverse temperatures $\hat\beta_k$. A symmetric proposal to swap two physical configurations, accepted with $A_x$ in \eqref{eq:main-exchange}, is reversible with respect to $\prod_k\nu_{G,k}$. The analogous auxiliary swap with acceptance $A_s$ is reversible with respect to $\prod_kP_{G,k}$.
\end{siproposition}

\begin{proof}
For an auxiliary swap between levels $i$ and $j$, the ratio of the swapped and original product densities is
\[
 \frac{P_{G,i}(s^j)P_{G,j}(s^i)}{P_{G,i}(s^i)P_{G,j}(s^j)}
 =\exp\!\left[(\hat\beta_i-\hat\beta_j)(G(s^i)-G(s^j))\right].
\]

For a physical swap, write $G_i=G(s(x^i))$ and $G_j=G(s(x^j))$. The common physical factors $\varrho(x^i)\varrho(x^j)$ and the normalizers cancel, leaving
\begin{align*}
 \log\frac{\nu_{G,i}(x^j)\nu_{G,j}(x^i)}
              {\nu_{G,i}(x^i)\nu_{G,j}(x^j)}
 &=(\beta-\hat\beta_i)G_j+(\beta-\hat\beta_j)G_i\\
 &\quad-(\beta-\hat\beta_i)G_i-(\beta-\hat\beta_j)G_j\\
 &=(\hat\beta_i-\hat\beta_j)(G_i-G_j).
\end{align*}
Thus the two log ratios are
\begin{equation}\label{eq:physical-exchange}
 \ell_s=(\hat\beta_i-\hat\beta_j)[G(s^i)-G(s^j)],
 \qquad
 \ell_x=(\hat\beta_i-\hat\beta_j)[G_i-G_j].
\end{equation}

In either case, let $T$ denote the swap and $\Pi$ the product density. Since $T^2z=z$, the Metropolis acceptance $A(z)=\min\{1,\Pi(Tz)/\Pi(z)\}$ satisfies
$\Pi(z)A(z)=\min\{\Pi(z),\Pi(Tz)\}=\Pi(Tz)A(Tz)$.
Together with the symmetric proposal, this proves detailed balance. With equal masses and a common thermostat, the Maxwell factors also cancel if momenta are exchanged. Leaving momenta in place preserves the same independent Maxwell law. Composing pair-exchange kernels therefore preserves $\Pi$. A joint symmetric swap of both populations instead uses the sum $\ell_x+\ell_s$ as its log acceptance ratio.
\end{proof}

\subsection{Convergence of the coupled population dynamics}\label{app:convergence}\label{sec:convergence}

We now consider the overdamped population dynamics with continuous parameter descent and no replica exchange. The aim is to control the objective error together with the lag of the sampling distributions behind their changing targets.

First suppose that all levels satisfy $0<\alpha_k<1$, with $c_k>0$ and $\sum_kc_k=1$. We treat the physical endpoint below. It is convenient to use dimensionless quantities,
\[
 g_\theta=\beta G_\theta,\qquad
 F(\theta)=\beta\mathcal L_{\mathrm{VESPER}}[G_\theta],\qquad
 f_\theta(s)=\nabla_\theta g_\theta(s),
\]
so the instantaneous targets are
\begin{equation}\label{eq:convergence-targets}
 P_k^\theta(s)=\frac{e^{-\alpha_kg_\theta(s)}}{A_k(\theta)},
 \qquad
 \nu_k^\theta(x)=\frac{\varrho(x)e^{(1-\alpha_k)g_\theta(s(x))}}{B_k(\theta)}.
\end{equation}
Let $\rho_{k,t}$ and $\nu_{k,t}$ denote the evolving auxiliary and physical laws. Suppressing the time subscript inside expectations, their equations are
\begin{equation}\label{eq:convergence-fp}
 \begin{aligned}
 \partial_t\rho_k
 &=d_{s,k}\nabla_s\cdot\left(\rho_k\nabla_s\log\frac{\rho_k}{P_k^{\theta_t}}\right),
 &d_{s,k}&=(\beta\gamma_{s,k})^{-1},\\
 \partial_t\nu_k
 &=d_{x,k}\nabla_x\cdot\left(\nu_k\nabla_x\log\frac{\nu_k}{\nu_k^{\theta_t}}\right),
 &d_{x,k}&=(\beta\gamma_{x,k})^{-1},\\
 \dot\theta_t
 &=-\eta\sum_kc_k\left(
 \mathbb E_{\nu_k}[f_{\theta_t}\circ s]-\mathbb E_{\rho_k}f_{\theta_t}\right).
 \end{aligned}
\end{equation}
Here $\eta=(\beta\gamma_\theta)^{-1}$ matches the normalization in \eqref{eq:sgda}. The physical drift is available directly from
$\nabla_x\log\nu_k^\theta=-\beta\nabla_xU+(1-\alpha_k)\nabla_x(g_\theta\circ s)$.

Measure the sampling lag by
\begin{equation}\label{eq:tracking-entropy}
 H_t=\sum_kc_k\left\{
 \frac{\operatorname{KL}(\rho_{k,t}\|P_k^{\theta_t})}{\alpha_k}
 +\frac{\operatorname{KL}(\nu_{k,t}\|\nu_k^{\theta_t})}{1-\alpha_k}
 \right\}.
\end{equation}
These weights will cancel the tempering factors when we differentiate the relative entropies.

Let $F_* =\min_\theta F(\theta)$ be finite and attained. We make the following assumptions on an open neighborhood $\mathcal O$ of the initial sublevel $\{\theta:F(\theta)\le F(\theta_0)+H_0\}$:
\begin{enumerate}[label=(C\arabic*)]
 \item The normalizers are finite, $g_\theta$ is differentiable in $\theta$ and smooth enough in space for \eqref{eq:convergence-fp}, and
 $\sup_{\theta\in\mathcal O,s}\|f_\theta(s)\|\le B<\infty$.
 Differentiation under the integrals is justified.
 \item The evolution is well posed from initial laws with finite relative entropy. Entropy differentiation and integration by parts are valid, with periodic or zero-flux boundaries, or sufficient decay on unbounded domains.
 \item The instantaneous laws satisfy uniform log-Sobolev inequalities on their respective spaces:
 \[
 I(\zeta\|P_k^\theta)\ge2\lambda_{s,k}\operatorname{KL}(\zeta\|P_k^\theta),
 \qquad
 I(\zeta\|\nu_k^\theta)\ge2\lambda_{x,k}\operatorname{KL}(\zeta\|\nu_k^\theta),
 \]
 where $I(\zeta\|\pi)=\int\|\nabla\log(\zeta/\pi)\|^2\,\dd\zeta$.
 With constants positive and uniform on $\mathcal O$, set
 $a=\min_k\{d_{s,k}\lambda_{s,k},d_{x,k}\lambda_{x,k}\}>0$.
 \item The objective satisfies the Polyak--\L{}ojasiewicz inequality
 $\|\nabla F(\theta)\|^2\ge2\mu(F(\theta)-F_*)$
 on $\mathcal O$, for some $\mu>0$.
\end{enumerate}
In (C3), the physical inequality is required for the full microscopic law. Uniform constants follow, for example, from the Holley--Stroock perturbation lemma when the model gives a uniformly bounded tilt of a reference law satisfying a log-Sobolev inequality \citep{holleystroock1987}.

\begin{sitheorem}[Convergence of the two-population ladder]\label{thm:convergence}
Under (C1)--(C4), for $0<\eta<a/B^2$, the evolution in \eqref{eq:convergence-fp} obeys
\begin{equation}\label{eq:main-convergence}
 F(\theta_t)-F_*+H_t
 \le e^{-\kappa t}\bigl[F(\theta_0)-F_*+H_0\bigr],
 \qquad \kappa=2\min\{\eta\mu,a-\eta B^2\}.
\end{equation}
Thus the objective error and the weighted sampling lag decay together.
\end{sitheorem}

\begin{proof}
Let $e$ denote the error in the population gradient, and write
\begin{align*}
 H_{s,k}&=\operatorname{KL}(\rho_k\|P_k^{\theta_t}),\qquad
 H_{x,k}=\operatorname{KL}(\nu_k\|\nu_k^{\theta_t}),\\
 e_k&=(\mathbb E_{\nu_k}-\mathbb E_{\nu_k^{\theta_t}})[f_{\theta_t}\circ s]
       -(\mathbb E_{\rho_k}-\mathbb E_{P_k^{\theta_t}})f_{\theta_t},
 \qquad e=\sum_kc_ke_k.
\end{align*}
By \eqref{eq:ladder-gradient}, the parameter equation is
$\dot\theta=-\eta(\nabla F+e)$.

To differentiate $H$, first note that normalization of the target densities gives
\[
 \nabla_\theta\log P_k^\theta
 =-\alpha_k(f_\theta-\mathbb E_{P_k^\theta}f_\theta),\qquad
 \nabla_\theta\log\nu_k^\theta
 =(1-\alpha_k)(f_\theta\circ s-\mathbb E_{\nu_k^\theta}[f_\theta\circ s]).
\]
Integration by parts in \eqref{eq:convergence-fp} gives
\begin{align}
 \dot H_{s,k}
 &=-d_{s,k}I(\rho_k\|P_k^{\theta_t})
  +\alpha_k\dot\theta^\top
       (\mathbb E_{\rho_k}-\mathbb E_{P_k^{\theta_t}})f_{\theta_t},
 \label{eq:entropy-s}\\
 \dot H_{x,k}
 &=-d_{x,k}I(\nu_k\|\nu_k^{\theta_t})
  -(1-\alpha_k)\dot\theta^\top
       (\mathbb E_{\nu_k}-\mathbb E_{\nu_k^{\theta_t}})[f_{\theta_t}\circ s].
 \label{eq:entropy-x}
\end{align}
Multiplying by $c_k/\alpha_k$ and $c_k/(1-\alpha_k)$, respectively, and summing over levels combines the moving-target terms into $-\dot\theta^\top e$. The log-Sobolev inequalities in (C3) therefore give
\begin{equation}\label{eq:entropy-tracking-bound}
 \dot H\le-2aH-\dot\theta^\top e.
\end{equation}

The gradient error is controlled by the same entropies. For a vector-valued function $f$ with $\|f\|\le B$, applying Pinsker's inequality to unit scalar projections gives
$\|(\mathbb E_\zeta-\mathbb E_\pi)f\|\le\sqrt2B\sqrt{\operatorname{KL}(\zeta\|\pi)}$.
Applying this to both populations and then using weighted Cauchy--Schwarz yields
\begin{align}
 \|e\|^2
 &\le2B^2\left[\sum_kc_k(\sqrt{H_{s,k}}+\sqrt{H_{x,k}})\right]^2\nonumber\\
 &\le2B^2
 \left[\sum_kc_k\{\alpha_k+(1-\alpha_k)\}\right]
 \left[\sum_kc_k\left\{\frac{H_{s,k}}{\alpha_k}
                       +\frac{H_{x,k}}{1-\alpha_k}\right\}\right]
 =2B^2H.
 \label{eq:si-gradient-error}
\end{align}

Now let $V=F-F_*+H$. Combining \eqref{eq:entropy-tracking-bound} with the parameter equation cancels the mixed gradient terms:
\begin{align*}
 \dot V
 &\le(\nabla F-e)^\top\dot\theta-2aH\\
 &=-\eta(\nabla F-e)^\top(\nabla F+e)-2aH\\
 &=-\eta\|\nabla F\|^2+\eta\|e\|^2-2aH\\
 &\le-\eta\|\nabla F\|^2-2(a-\eta B^2)H
 \le-\kappa V.
\end{align*}
The last inequality uses (C4), and Gronwall's inequality gives \eqref{eq:main-convergence}. This estimate also gives $F(\theta_t)\le F_*+V_t\le F(\theta_0)+H_0$. Applying the argument up to a possible first exit from $\mathcal O$ therefore rules out such an exit and proves the estimate for all time.
\end{proof}

\paragraph{The physical endpoint.}
The same argument covers a ladder with $\alpha_K=1$, where the last physical replica has the fixed target $\nu_K^\theta=\varrho$. The gradient formula in \eqref{eq:ladder-gradient} still holds under the endpoint moment conditions of \cref{prop:kl-control}, as shown in \cref{prop:ladder-si}. To adapt the entropy $H$, keep the terms for $k<K$ and replace the last pair by
\[
 c_K\{\operatorname{KL}(\rho_K\|P_K^\theta)
       +w\operatorname{KL}(\nu_K\|\varrho)\},\qquad w>0.
\]
Write $H'$ for this modified entropy and use (C1)--(C4) on an open neighborhood of $\{\theta:F(\theta)\le F(\theta_0)+H'_0\}$, with the physical log-Sobolev inequality at level $K$ imposed on $\varrho$.

Since $\varrho$ does not depend on $\theta$, its relative entropy has no moving-target term. Thus, with
$e_{x,K}=(\mathbb E_{\nu_K}-\mathbb E_{\varrho})[f_{\theta}\circ s]$, the entropy estimate becomes
\[
 \dot H'\le-2aH'-\dot\theta^\top(e-c_Ke_{x,K}).
\]
Pinsker's inequality and the same weighted Cauchy--Schwarz estimate give
\[
 \|e\|^2\le2B^2(1+c_K/w)H',\qquad
 c_K^2\|e_{x,K}\|^2\le2B^2(c_K/w)H'.
\]
For $V'=F-F_*+H'$, the mixed terms therefore cancel as before, except for the contribution from the last physical replica:
\begin{align*}
 \dot V'
 &\le-\eta\|\nabla F\|^2+\eta\|e\|^2
      -\eta c_K(\nabla F+e)^\top e_{x,K}-2aH'\\
 &\le-\tfrac{\eta}{2}\|\nabla F\|^2
      -\bigl[2a-\eta B^2(3+5c_K/w)\bigr]H'.
\end{align*}
Here we used Young's inequality in the form
$|c_K(\nabla F+e)^\top e_{x,K}|\le\frac12\|\nabla F\|^2+\frac12\|e\|^2+c_K^2\|e_{x,K}\|^2$.
For $0<\eta<2a/[B^2(3+5c_K/w)]$, (C4) then gives
\[
 V'_t\le e^{-\kappa' t}V'_0,\qquad
 \kappa'=\min\{\eta\mu,\,2a-\eta B^2(3+5c_K/w)\}.
\]
As in the proof above, $F(\theta_t)\le F(\theta_0)+H'_0$ keeps the trajectory in the region where the assumptions hold. Without (C4), integrating the bound on $\dot V'$ gives the time-averaged estimate in \eqref{eq:si-stationarity}, with $2V'_0/(\eta T)$ on the right.

\begin{sicorollary}[Time-averaged stationarity]\label{cor:convergence-nonconvex}
Under (C1)--(C3) and $0<\eta<a/B^2$, for every $T>0$,
\begin{equation}\label{eq:si-stationarity}
 \frac1T\int_0^T\|\nabla F(\theta_t)\|^2\,\dd t
 \le\frac{F(\theta_0)-F_*+H_0}{\eta T}.
\end{equation}
\end{sicorollary}
\begin{proof}
The proof above gives $\dot V\le-\eta\|\nabla F\|^2-2(a-\eta B^2)H$ before invoking (C4). Integrating over $[0,T]$ and using $V_T\ge0$ proves the bound.
\end{proof}

\subsection{General reference measures}\label{app:reference-measure}

The same variational argument applies to a general reference measure. Let $r$ be sigma-finite and let the probability measure $p$ have density $f=\dd p/\dd r>0$ $r$-almost everywhere. For $r=\dd s$, this $f$ is the physical CV density $p(s)$. Writing $g=\beta G$, define
\begin{equation}\label{eq:dimensionless-objective}
 \ell_\alpha[g]
 =\frac1\alpha\log\int e^{-\alpha g}\,\dd r
  +\frac1{1-\alpha}\log\int e^{(1-\alpha)g}\,\dd p.
\end{equation}
For the physical Lebesgue reference, $\ell_\alpha[\beta G]=\beta\mathcal L_{\hat\beta}[G]$. When $Z_g=\int e^{-g}\,\dd r<\infty$, the normalized residual model is
\begin{equation}\label{eq:residual}
 q_g(\dd y)=Z_g^{-1}e^{-g(y)}r(\dd y).
\end{equation}

\begin{siproposition}[Reference-measure minimizer]\label{prop:reference-minimizer}
For $0<\alpha<1$ with $M(\alpha)=\int f^\alpha\,\dd r<\infty$, the functional in \eqref{eq:dimensionless-objective}, restricted to $g$ for which both integrals are finite, is minimized exactly at $g_\star=-\log f+C$. At the minimum, the two tilted laws coincide with $f^\alpha\dd r/M(\alpha)$.
\end{siproposition}
\begin{proof}
As in the proof of \cref{thm:consistency}, H\"older's inequality gives
\[
 M(\alpha)=\int(e^{-\alpha g})^{1-\alpha}
                    (f e^{(1-\alpha)g})^\alpha\,\dd r
 \le\left(\int e^{-\alpha g}\,\dd r\right)^{1-\alpha}
      \left(\int e^{(1-\alpha)g}\,\dd p\right)^\alpha.
\]
Equality requires the two factors to be proportional, which is equivalent to $g=-\log f+C$. At this choice, both normalized densities relative to $r$ equal $f^\alpha/M(\alpha)$.
\end{proof}

\subsection{Positive-weight effective sample size}

\begin{siproposition}[Population normalized ESS]\label{prop:bridge}
At the population optimum, the fixed-data weights are proportional to $f^{\alpha-1}$. If $M(2\alpha-1)<\infty$, their population normalized effective sample size is
\begin{equation}\label{eq:ess}
 \frac{(\mathbb E_pw)^2}{\mathbb E_pw^2}
 =\frac{M(\alpha)^2}{M(2\alpha-1)}.
\end{equation}
For independent $y_j\sim p$ with $w_j=w(y_j)$, the empirical quantity $(\sum_jw_j)^2/[n\sum_jw_j^2]$ converges to this value almost surely. At $\alpha=1/2$, the denominator is $M(0)=r(\operatorname{supp}p)$.
\end{siproposition}
\begin{proof}
The scale of the weights cancels in the ratio, so take $w=f^{\alpha-1}$. Since $\dd p=f\dd r$, we have $\mathbb E_pw=\int f^\alpha\,\dd r=M(\alpha)$ and $\mathbb E_pw^2=\int f^{2\alpha-1}\,\dd r=M(2\alpha-1)$, proving \eqref{eq:ess}. The assumed second moment allows the strong law of large numbers to be applied to both empirical moments, giving the almost-sure limit.
\end{proof}

\subsection{Learning from existing data}\label{sec:data-learning}

When equilibrium configurations $x_j^*\sim\varrho$ are already available, their CV values $y_j=s(x_j^*)$ can replace the physical walkers. The physical expectation in \eqref{eq:physical-gradient} is then estimated using normalized importance weights
\begin{equation}\label{eq:main-data-weights}
 \bar w_j(\theta)
 =\frac{e^{(\beta-\hat\beta)G_\theta(y_j)}}
        {\sum_\ell e^{(\beta-\hat\beta)G_\theta(y_\ell)}}.
\end{equation}

This leads to the following weighted-data consistency result.

\begin{siproposition}[Weighted-data gradient]\label{lem:offline}
Fix $\theta$ and let $x_j^*$ be independent samples from $\varrho$, with $y_j=s(x_j^*)$. Suppose $B_{G_\theta}<\infty$ and
$\mathbb E_p[e^{(\beta-\hat\beta)G_\theta}\|\nabla_\theta G_\theta\|]<\infty$.
Then the weights in \eqref{eq:main-data-weights} satisfy
\begin{equation}\label{eq:si-data-consistency}
 \sum_{j=1}^n\bar w_j(\theta)\nabla_\theta G_\theta(y_j)
 \longrightarrow\mathbb E_{\nu_{G_\theta}}\nabla_\theta G_\theta\circ s
 \quad\text{almost surely as }n\to\infty.
\end{equation}
Moreover, the corresponding weighted population objective has minimizer $G_\star+C$.
\end{siproposition}

\begin{proof}
Using the CV marginal in \eqref{eq:physical-cv-law}, the required expectation can be written as
\[
 \mathbb E_{\nu_{G_\theta}}\nabla_\theta G_\theta\circ s
 =\frac{\mathbb E_p[e^{(\beta-\hat\beta)G_\theta}\nabla_\theta G_\theta]}
        {\mathbb E_p[e^{(\beta-\hat\beta)G_\theta}]}.
\]
The assumed moments allow the strong law of large numbers to be applied to both sample averages. Their ratio converges almost surely because the limiting denominator is positive, proving \eqref{eq:si-data-consistency}.

Together with the auxiliary expectation, this yields the gradient in \eqref{eq:physical-gradient} of the original population objective, whose minimizer is $G_\star+C$ by \cref{thm:consistency}. The exponential weighting is essential: using unweighted samples from $p$ would instead give the functional stationarity condition $P_G=p$, equivalently $G=G_\star/\alpha+C$.
\end{proof}

The auxiliary walkers still sample $P_{G_\theta}$, so the resulting update is a weighted-data version of the same learning rule. With $g=\beta G$ and $\gamma=\alpha-1$, the objective coincides with the PS-CD/$\gamma$-cross-entropy objective continued to negative $\gamma$ \citep{yu2021pseudospherical,fujisawa2008robust}, with the R\'enyi variational representation described in \cref{app:related} \citep{birrell2021renyi}.

The same ratio argument also yields within-population sample-reuse weights. If a batch was generated at $G_0$, then the current-to-generating density ratios are proportional to
\begin{equation}\label{eq:si-reuse-weights}
 w^{(s)}=e^{-\hat\beta(G-G_0)},\qquad
 w^{(x)}=e^{(\beta-\hat\beta)(G-G_0)\circ s}.
\end{equation}
Normalization within each pool removes the unknown partition functions:
\[
\mathbb E_{P_{G_0}}[w^{(s)}f]=\frac{A_G}{A_{G_0}}\,\mathbb E_{P_G}f,\qquad
\mathbb E_{\nu_{G_0}}[w^{(x)}f\circ s]=\frac{B_G}{B_{G_0}}\,\mathbb E_{\nu_G}f\circ s.
\]
Hence, if $G$ and $G_0$ are admissible, $f$ is integrable under the current law, and the batch satisfies a law of large numbers under its generating law, the corresponding self-normalized estimators are consistent. Their weight concentration is monitored by the ESS criterion described in \cref{app:extra}.

The single-level update also admits a scalar-objective interpretation in self-consistent well-tempered VES. For $V_\theta=-(1-\alpha)G_\theta$ and the instantaneous target $t_\theta=P_{G_\theta}$, differentiating the VES functional while holding $t$ fixed gives
\begin{equation}\label{eq:ves-correspondence}
 \left.\nabla_\theta\Omega[V_\theta;t]\right|_{t=t_\theta}
 =(1-\alpha)\nabla_\theta\mathcal L_{\hat\beta}[G_\theta].
\end{equation}
The target convention and the recursive neural VES update are detailed in \cref{app:ves-correspondence} \citep{Valsson_2014,valsson2015welltempered,bonatiNeuralNetworksbasedVariationally2019a}.
\section{Physical systems: models and protocols}\label{app:physics}
This appendix gives the model, force field, collective-variable and network settings behind \cref{sec:experiments}, together with the trust region, readout and gauge conventions used for every number reported there.

Unless stated otherwise, reported estimates are means $\pm$ sample standard deviations across independent runs. Errors are computed for each run before averaging. Free-energy maps formed by averaging model predictions before marginalization are identified explicitly. Molecular times are per physical replica, and aggregate MD is summed over replicas.

\subsection{Alanine dipeptide}\label{app:physics-adp}
\paragraph{System and CVs.}
Explicit-solvent alanine dipeptide, AMBER99SB \citep{hornakComparisonMultiple2006} with TIP3P water \citep{jorgensenComparisonSimple1983} and particle-mesh Ewald electrostatics, 298.15 K, 2 fs timestep, bonds to hydrogen constrained. The CVs are the backbone dihedrals $\phi,\psi$, both periodic.

\paragraph{Model and training.}
$G_\theta$ is a SIREN \citep{sitzmannImplicitNeural2020} with one hidden layer of width $32$, $\omega_0=16$ and a periodic input embedding. Each replica carries $10^4$ auxiliary walkers with $\gamma_s=10$ and $\gamma_x=100$. Training uses AdamW \citep{loshchilovDecoupledWeight2019} at learning rate $3\times10^{-4}$, up to $80$ inner iterations per epoch, and epochs of $1000$ MD steps, $50$ per checkpoint ($100$ ps). The single replica uses $\alpha=0.01$. Both four-level ladders use $\alpha=(0.1,0.05,0.02,0.01)$ with equal loss weights. L-RE attempts exchange of physical configurations between adjacent levels every epoch. L-NRE uses the same ladder without exchange. The single replica starts from the same configurations as the OPES runs.

\paragraph{Reference and analysis.}
The reference surface is the free energy from 100~ns of adaptive-biasing-force sampling by \citet{zubietarico2025sobolev}, on a $100\times100$ periodic grid. \Cref{fig:res-adp} reports mean-aligned RMSE over this full grid. OPES and TAMD are scored only on cells where their native free-energy estimators are finite. VESPER predictions use parameter averages over the trailing eight available checkpoints, or all available checkpoints at earlier times. RMSE curves average the individual-run errors. The displayed surfaces average the mean-aligned predictions and shift the resulting minimum to zero. Physical points overlaid on the maps are drawn from the last nanosecond and auxiliary points from the final saved populations, balanced across runs and ladder levels.

\subsection{Alanine tetrapeptide}\label{app:physics-atp}
\paragraph{System and CVs.}
Gas-phase Ace-(Ala)$_3$-NMe, AMBER99SB \citep{hornakComparisonMultiple2006} without cutoff, 298.15 K, 2 fs timestep, bonds to hydrogen constrained. The three-CV model biases $(\phi_1,\phi_2,\phi_3)$ and the six-CV model adds $\psi_1$, $\psi_2$ and $\psi_3$.

\paragraph{Model and training.}
$G_\theta$ is a SIREN with hidden widths $(14,4)$, $\omega_0=16$ and a periodic input embedding, so the two models differ only in the input layer ($163$ and $247$ parameters). Each replica carries $2\times10^4$ auxiliary walkers with $\gamma_s=\gamma_x=100$. Training uses AdamW at $3\times10^{-4}$, up to $80$ inner iterations, and epochs of $1000$ steps, $100$ per checkpoint ($200$ ps). The single replica uses $\alpha=0.01$, and the ladder arms are those of ADP. Both models train at a constant step size throughout. Each model is read as the parameter-SWA10 average over checkpoints from 8.2--10.0 ns.

\paragraph{Readout and analysis.}
Temporal correlations compare trailing ten-checkpoint parameter averages ending at 5 and 10~ns, using windows of 3.2--5.0 and 8.2--10.0~ns, respectively. Pearson correlations are calculated on 262,144 evaluation points pooled from four scrambled Sobol sets spanning the full six-dimensional CV domain. Reported values are the mean and sample standard deviation across three independent runs.

\Cref{fig:res-atp} shows physical Boltzmann marginals at 298.15~K. The full physical free-energy predictions from the three independent runs of the four-level exchange ladder are averaged before Boltzmann integration and marginalization. The retained $(\phi_1,\phi_2)$, $(\phi_1,\phi_3)$ and $(\phi_2,\phi_3)$ planes are complete $64\times64$ grids. Six-CV maps integrate omitted $\psi$ coordinates on 32-point axes and the remaining omitted $\phi$ on its 64-point grid, while three-CV maps integrate the omitted $\phi$ on a 256-point grid. Display surfaces are shifted to zero minimum and use a common 0--120 kJ\,mol$^{-1}$ scale with 10 kJ\,mol$^{-1}$ contours. The three-to-six-CV squared-Pearson correlations over all plane cells are $0.951$, $0.895$ and $0.890$.

\subsection{Chignolin}\label{app:physics-chig}
\paragraph{System and CVs.}
CLN025 \citep{hondaCrystalStructure2008} in explicit solvent, ff14SB \citep{maierff14SB2015} with TIP3P \citep{jorgensenComparisonSimple1983}, 340 K, 2 fs timestep, with its seventeen backbone dihedrals as CVs. Structural deviation is measured by C$\alpha$ root-mean-square deviation (RMSD) after Kabsch superposition \citep{kabschSolutionBest1976}.

\paragraph{Model and training.}
$G_\theta$ is a SIREN with hidden widths $(48,24)$, $\omega_0=1$ and a periodic input embedding ($2{,}881$ parameters). Twelve replicas span a geometric ladder from $\alpha=1$ to $\alpha=1/18$, every level exchanging physical configurations with its neighbors every epoch. Each carries $10^5$ auxiliary walkers. Training uses AdamW at learning rate $10^{-4}$ with a gradient clip of $10\,k_BT$, up to $100$ inner iterations and epochs of $5000$ steps, $20$ per checkpoint ($200$ ps), at constant step throughout. Four independent runs extend to $26.2$ ns per replica, and the readout is the final checkpoint.

\paragraph{Readout and analysis.}
Frames are saved every $5$ ps, $5{,}240$ per level and run. Full-run occupancy fractions and exchange acceptance pool the four runs over $26.2$~ns. Every reported unfolding and refolding event falls within the first $25$~ns. Folded and unfolded are defined as C$\alpha$ RMSD below $0.20$~nm and above $0.40$~nm. An unfolding or refolding event requires the new state to persist for $20$~ps and is counted along each configuration's identity. This dwell criterion is not applied to the framewise occupancy estimates.

The time-resolved occupancies in \cref{fig:res-chig}A,B use non-overlapping $0.2$~ns bins, each containing $40$ saved frames per run. At each saved time, configurations are assigned to their current tempering levels before RMSD classification. For panel B, the within-bin folded fractions at each level are averaged equally across the four runs. For panel A, each run's per-level fractions are summed to give the time-averaged numbers of levels in each state. The lines show their mean and the bands their minimum-to-maximum range across runs. Both panels display bins in the first $25$~ns. The reported $99.3\%$ unbiased-level occupancy uses the full $26.2$~ns trajectory.

The learned surfaces are estimated from $25{,}000$ auxiliary walkers of the unbiased level drawn at random from each run at the readout ($100{,}000$ in all), and the MD frames shown in \cref{fig:res-chig}C come from the last nanosecond. For the radius-of-gyration and RMSD map, each walker's dihedrals are placed on a backbone with the bond lengths, bond angles, peptide $\omega$ and Pro4 $\phi$ of the native structure, and $F=-k_BT\ln p$ ($k_BT=2.83$ kJ\,mol$^{-1}$), with $p$ an adaptive Gaussian kernel density estimate. The dihedral maps are product von Mises kernel density estimates, compared after mean alignment with the same estimate on the unbiased level's MD frames where the MD free energy is below $10$ kJ\,mol$^{-1}$. Over $(\phi_2,\psi_2)$, $(\psi_3,\psi_4)$ and $(\phi_5,\psi_5)$, the learned minimum lies $0.42$, $0.07$ and $0.19$ rad from the MD minimum, at points where the MD free energy is $1.3$, $0.2$ and $0.5$ kJ\,mol$^{-1}$ above its minimum, with RMS differences of $1.9$, $2.2$ and $2.0$ kJ\,mol$^{-1}$. Accepted exchanges permute configurations between levels, so every configuration keeps its identity (\cref{app:method-exchange}).

\subsection{Catalytic methane dehydrogenation}\label{app:physics-cat}
\paragraph{System and CVs.}
Methane on stepped Ni(511) at $700$ K uses an Allegro machine-learned interatomic potential \citep{musaelianLearningLocalEquivariant2023}, deployed through the NequIP toolchain \citep{batznerE3equivariantGraph2022} and driven by OpenMM \citep{eastmanOpenMM72017}. The CVs are the two coordination numbers of \citet{jinFreeEnergyLandscapesSurface2025} and the periodic fractional coordinate $S_1$ of the carbon atom along the first surface lattice vector. The system has $85$ atoms, $25$ held fixed, in a triclinic cell. A soft restraint keeps the system in the CV ranges: $S_1\in[0,1]$, $\mathrm{CNi}\in[0.05,3.02]$, and $\mathrm{CH}\in[2.95,3.98]$.

\paragraph{Model and training.}
$G_\theta$ is a SIREN with hidden widths $(14,4)$ and $\omega_0=16$. One replica at $\alpha=0.05$ carries $6.5\times10^4$ auxiliary walkers, using AdamW at learning rate $10^{-4}$, with up to $100$ inner iterations and epochs of $5000$ steps, $20$ per checkpoint, at a $0.25$ fs timestep, so one checkpoint is $25$ ps. Training uses a gradient clip of $50\,k_BT$, $\gamma_s=100$, $\gamma_x=10$, Langevin friction $9.82$ ps$^{-1}$ and a constant step. The reweighted inner step uses the full trust region of \cref{app:physics-trust}. Three independent runs extend to checkpoint 300. Each checkpoint-300 physical free-energy prediction uses deployed parameter EMA decay $0.9$.

\subsection{Catalysis free-energy projections}\label{app:submission-applications}
With $S_1$ the periodic fractional coordinate and $y,z$ the C--Ni and C--H coordination numbers, the two-dimensional free energies shown in \cref{fig:res-cat} are Boltzmann marginals of the equal-weight mean of the three gauge-aligned checkpoint-300 EMA predictions,
\[
 F_R(q_R)=-k_BT\log\int_{D_O(q_R)}\exp[-G_{\rm mean}(q_R,q_O)/(k_BT)]\,\mathrm dq_O,
\]
at $T=700$ K, with ordinary measure in the original CV coordinates and integration box $D=S_1[0,1]\times y[0.02,3]\times z[2.90,3.94]$, while the visible route uses $S_1[0.02,0.98]\times y[0.10,2.95]\times z[2.97,3.94]$. Each map is shifted to its own minimum for display. The route is initialized by a string-method calculation on the differentiable mean surface and refined by constrained gradient flow, resolving contacts with the route bounds. This yields a locally resolved route within the stated box, but global path optimality is not established. The physisorbed representative is constrained at the lower C--Ni route bound, and the fitted saddle is at $S_1\approx0.65$, near the step-edge atoms at $S_1\approx0.70$. The route's $G(3)-G(1)=1.12$ eV is a bounded surface difference. Its $0.25$ eV uncertainty is the sample standard deviation of the individual model differences evaluated at the same reactant and saddle points on the mean-surface route.
\subsection{Ladder configuration by system}\label{app:ladder-config}
\Cref{tab:ladder-config} lists the levels of every molecular system. Each level is one physical trajectory. The auxiliary walkers enter only the training loss. The single-level runs of ADP and ATP use $\alpha=0.01$, and ADP also has a single-level run at $\alpha=1$.

\begin{table}[t]
\centering
\small
\caption{Levels per molecular system: tempering exponents $\alpha$, number of levels $K$, and walkers per level.}
\label{tab:ladder-config}
\begin{tabular}{@{}llrrr@{}}
\toprule
System & $\alpha$ & $K$ & Physical & Auxiliary \\
\midrule
ADP & $0.1,\,0.05,\,0.02,\,0.01$ & 4 & 1 & $10^4$ \\
ATP, $D=3,\,6$ & $0.1,\,0.05,\,0.02,\,0.01$ & 4 & 1 & $2\times10^4$ \\
Methane/Ni(511) & $0.05$ & 1 & 1 & $6.5\times10^4$ \\
Chignolin & $18^{-k/11}$, $k=0,\dots,11$ & 12 & 1 & $10^5$ \\
\bottomrule
\end{tabular}
\end{table}

\subsection{Reweighted inner steps and the step-size controller}\label{app:physics-trust}
Within an epoch the walker samples generated under $G_{\theta_0}$ are frozen while several parameter steps are taken and reweighted by self-normalized importance sampling at every step (\cref{sec:algorithm}). A trust region bounds the log-weights: a soft threshold $\delta=5$ on their $95$th percentile, a hard threshold $\delta_{\mathrm{hard}}=8$ on their maximum, and a floor of $0.2$ on the effective sample size of the pool. Exceeding the hard threshold or falling below the floor ends the epoch and forces fresh trajectory data. Exceeding the soft threshold multiplies the parameter step by $1/q^2$ with $q=\lVert\log w\rVert_{95}/\delta$. It is used on ADP, ATP and catalysis. Chignolin trains on fresh samples only.\label{app:physics-controller}

On ADP, the inner-iteration budget and learning rate are governed by a sample-based controller rather than a predefined schedule. The controller latches a decay phase when cumulative relative parameter change across checkpoints exceeds a budget of $160\%$, then multiplies the learning rate by $0.95$ and the inner-iteration count by $0.7$ per checkpoint, flooring the count at $5$. Once the count reaches its floor and the learning rate is below $5\%$ of its initial value, a final phase continues the decay to a floor of $0.3\%$ and deploys an exponential moving average (EMA) of parameters with decay $0.99$ per epoch. ATP uses constant-step training.

\subsection{Readout, optimizer settings and RMSE gauge}\label{app:physics-gauge}
Models use system-specific parameter readouts. ATP predictions use a uniform average of the last ten checkpoints, and ADP uses the last eight available checkpoints. At the reported horizon these ADP parameters have already undergone EMA smoothing. These are the Polyak--Ruppert or stochastic-weight-averaging (SWA) readouts \citep{polyakAcceleration1992,ruppertEfficient1988,izmailovAveragingWeights2018}. Catalysis uses the deployed EMA readout with decay $0.9$ at checkpoint 300, and chignolin reports its final checkpoint ($26.2$~ns per replica). AdamW \citep{loshchilovDecoupledWeight2019} uses decoupled weight decay $w=0.01$ in every run, and the peptide runs apply the cautious mask of \citet{liangCautiousOptimizers2024}, which zeroes coordinates whose Adam step is not descent-aligned with the current gradient and rescales the rest.

RMSE is reported after removing the additive constant. \emph{Mean alignment}, used for ADP, removes the mean difference between the learned and reference surfaces on the scored region. Flat-surface baselines, where reported, use the same region and gauge. ATP dimensional-consistency comparisons use complete retained planes, squared-Pearson correlations and mean-aligned RMSEs after physical Boltzmann marginalization.

\subsection{Baseline settings}\label{app:comparison-plan}
The ADP and ATP comparisons in \cref{sec:adp,sec:atp} use the Hamiltonians and CVs specified above and a 2 fs timestep. The no-exchange control retains the ladder's architecture and population sizes. In ADP and ATP, auxiliary moves swap whole populations and were accepted in at most $0.5\%$ of epochs over 10~ns. Most accepted exchanges transport physical configurations. ADP VESPER single-replica, OPES and TAMD runs share initial configurations. Means and sample standard deviations use the trajectories that reach the reported horizon.

OPES uses bias factor 100 and barrier parameter $40$ kJ\,mol$^{-1}$ \citep{invernizziRethinkingMetadynamics2020}. Metadynamics deposits $1.2$ kJ\,mol$^{-1}$ hills every 500 steps \citep{laioEscapingFreeenergyMinima2002a}, and TAMD uses UFEDMM \citep{bajpaiInteroperableImplementationCollectivevariable2023}. Readouts are the OPES kernel density, the negative metadynamics hill sum, and the TAMD auxiliary-variable histogram at its own temperature. CFF \citep{sevgen2020cff} and its SIREN variant (S-CFF) \citep{zubietarico2025sobolev} use their saved native free-energy functions. ADP errors use the full-grid convention for methods with finite estimates and the native finite-estimator-cell convention for OPES and TAMD.
\paragraph{Neural VES.}\label{app:author-neural-ves}
We test this method using the released code \citep{bonatiNeuralNetworksbasedVariationally2019a}. The bias factor is 100, with 50 grid bins per periodic CV and ReLU widths $48$--$24$--$12$ for ADP.
Adam starts at $10^{-3}$. Parameter and target updates occur every 500 MD steps with the released system-specific learning-rate controllers. Errors use the native free-energy exports.

\subsection{Compute}\label{app:compute}
\Cref{tab:compute} lists the simulated time per replica at the readout and the resulting aggregate MD per run. Several runs continued beyond the readout, and the table counts MD up to it. The ADP baselines are OPES, metadynamics, TAMD, neural VES, CFF and S-CFF. VESPER, CFF and S-CFF ran on NVIDIA H100, H200, L40S and RTX PRO 6000 GPUs, and the other baselines on CPUs, so wall-clock times are not comparable across methods. For scale, one chignolin run takes about 100~h on one H100 GPU, which runs all twelve replicas, and one methane run takes about 140--180 GPU-hours, mostly on L40S GPUs.

\begin{table}[t]
\centering
\small
\setlength{\tabcolsep}{4pt}
\caption{Replicas $K$, simulated time per replica at the readout, and aggregate MD per run ($K$ times the time per replica).}
\label{tab:compute}
\begin{tabular}{@{}llrrrr@{}}
\toprule
System & Method & $K$ & ns/replica & aggregate ns & Runs \\
\midrule
ADP & VESPER, $\alpha=0.01$ or $1$ & 1 & 10 & 10 & 3 \\
ADP & VESPER ladders (L-RE, L-NRE) & 4 & 10 & 40 & 3 \\
ADP & Each baseline & 1 & 10 & 10 & 3 \\
ATP, $D=3,\,6$ & VESPER, $\alpha=0.01$ & 1 & 10 & 10 & 3 \\
ATP, $D=3,\,6$ & VESPER ladders (L-RE, L-NRE) & 4 & 10 & 40 & 3 \\
Methane/Ni(511) & VESPER, $\alpha=0.05$ & 1 & 7.5 & 7.5 & 3 \\
Chignolin & VESPER ladder, $\alpha=1$ to $1/18$ & 12 & 26.2 & 314.4 & 4 \\
\bottomrule
\end{tabular}
\end{table}

\subsection{Reported molecular results}\label{app:reported}
\Cref{tab:reported} collects the values quoted in \cref{sec:adp,sec:atp,sec:reactive,sec:chignolin}, with the readout behind each.

\begin{table}[t]
\centering
\footnotesize
\setlength{\tabcolsep}{3pt}
\caption{Values reported in the main text for the molecular systems. ADP: 10 ns per replica, SWA8 readout, three independent runs. ATP: 10 ns, SWA10, three independent runs, mean surfaces for the plane comparisons. Methane: 7.5 ns, EMA readout, three independent runs. Its $1.12\pm0.25$~eV entry is the route value on the mean surface with the sample SD evaluated at fixed route points. Chignolin: event counts use the first 25 ns per replica and the unbiased-level occupancy uses the full 26.2 ns trajectory, four independent runs. RMSEs are mean $\pm$ sample SD over runs.}
\label{tab:reported}
\begin{tabular}{@{}lll@{}}
\toprule
System & Quantity & Value \\
\midrule
ADP & RMSE, VESPER $\alpha=1$ (kJ\,mol$^{-1}$) & $13.27\pm0.26$ \\
ADP & RMSE, VESPER $\alpha=0.01$ & $2.19\pm0.02$ \\
ADP & RMSE, L-RE / L-NRE & $2.02\pm0.24$ / $2.05\pm0.09$ \\
ADP & RMSE, MetaD/VES/S-CFF & $2.61\pm0.22$/$4.72\pm0.25$/$2.04\pm0.51$ \\
ATP & $r^2$, 3-CV vs 6-CV marginals, three planes & $0.951$ / $0.895$ / $0.890$ \\
ATP & RMSE of the same planes (kJ\,mol$^{-1}$) & $5.5$ / $8.1$ / $7.5$ \\
ATP & $r$, 6-CV, 5 vs 10 ns, L-RE/L-NRE/single & $0.913{\pm}0.012$/$0.941{\pm}0.028$/$0.804{\pm}0.179$ \\
ATP & $r^2$ vs 3-CV on $(\phi_1,\phi_2)$, L-RE/L-NRE/single & $0.951$/$0.932$/$0.927$ \\
ATP & $r^2$ vs 3-CV on $(\phi_1,\phi_3)$, L-RE/L-NRE/single & $0.895$/$0.915$/$0.935$ \\
ATP & $r^2$ vs 3-CV on $(\phi_2,\phi_3)$, L-RE/L-NRE/single & $0.890$/$0.940$/$0.936$ \\
Methane/Ni(511) & $\Delta G$, physisorbed to transition region (eV) & $1.12\pm0.25$ \\
Chignolin & Unfolding / refolding events & $48$ / $9$ \\
Chignolin & Unbiased-level occupancy, full $26.2$ ns & $99.3\%$ \\
Chignolin & Exchange acceptance between adjacent levels & $30$--$93\%$ \\
Chignolin & Unfolded levels outnumber folded from & $3.9$ ns \\
Chignolin & $(\psi_3,\psi_4)$ minimum offset from unbiased MD & $0.07$ rad \\
\bottomrule
\end{tabular}
\end{table}

\section{Controlled landscapes}\label{app:controlled-overview}

\subsection{M\"uller--Brown protocol}\label{app:mb-ablation}
\paragraph{\Cref{fig:res-toy}A--C.}
The ground-truth energy is $0.1$ times the standard M\"uller--Brown potential at $\beta=1$ on the physical box $[-1.5,1]\times[-0.5,2]$, with the identity CV map. Seven arms share every setting except the tempering and, for the ladders, exchange: single levels at $\alpha=1$, $0.5$, $0.25$, $0.1$, $0.05$ with $1{,}000$ physical and $1{,}000$ auxiliary walkers, and a five-level ladder ($\alpha=1$, $0.5$, $0.25$, $0.1$, $0.05$; $200$ walkers per level and population; normalized level weights $0.2$) with and without exchange of physical and auxiliary walkers (every $100$ steps, $200$ attempted swaps). All physical walkers start within $\pm0.025$ of the deepest minimum, near $(-0.558,1.442)$, and the auxiliary walkers start uniformly over the box. Every arm of a run receives the same initial arrays and initial model. The network has width 32 and is trained with Adam at $3\times10^{-4}$. The dynamics use reduced timesteps $\Delta t_x=\Delta t_s=0.002$, unit frictions, force clips of $100$ and $500$, and reflecting boundaries. $40{,}000$ steps give $t=80$. Curves show the causal trailing-half parameter average: at each plotted time $T$ the raw parameters saved in $[T/2,T]$ are averaged, so no point uses a future state.

Errors are evaluated on a $241\times241$ grid over $[-1.5,1.5]\times[-0.5,2.5]$, which extends beyond the simulation box so that the high-energy rim is scored. The RMSE removes the additive constant and weights every cell equally. Because the rim energies are large it is dominated by the barrier and rim regions, which is why its scale is $237$--$252\,k_BT$ and why the most strongly tempered level is best on it. $D_{\mathrm{KL}}(\rho\,\|\,\rho_\theta)$ normalizes $\rho\propto e^{-F}$ and $\rho_\theta\propto e^{-G_\theta}$ on the same grid and is dominated by the basins. At $t=80$ (mean $\pm$ sample SD over three independent runs) the RMSE is $237.7\pm2.5$ ($\alpha=0.05$), $238.8\pm1.4$ (L-NRE), $239.0\pm1.1$ ($\alpha=0.1$), $240.0\pm1.2$ (L-RE), $243.1\pm2.0$ ($\alpha=1$), $243.7\pm0.4$ ($\alpha=0.25$) and $243.9\pm0.5$ ($\alpha=0.5$). The divergence is $0.010\pm0.008$ ($\alpha=0.5$), $0.010\pm0.007$ ($\alpha=0.25$), $0.017\pm0.013$ (L-RE), $0.019\pm0.013$ (L-NRE), $0.049\pm0.034$ ($\alpha=0.1$), $0.102\pm0.138$ ($\alpha=0.05$) and $0.82\pm1.14$ ($\alpha=1$). The physical walkers of the $\alpha=1$ arm remain concentrated in the starting basin ($355$ of the $360$ overlaid positions from $t=40$--$80$ lie nearest the deepest minimum), so its surface is unconstrained outside it. The maps in \cref{fig:res-toy}C are means of the final trailing-half averages after aligning the additive constants, each shifted to its own minimum. The walker overlays pool $40$ physical and $40$ auxiliary positions per run from each of $t=40$, $60$ and $80$.

\paragraph{Tempering and population sweeps.}
Separate sweeps at $t=320$ ($160{,}000$ steps) use four independent runs, uniform starts and a network of width 32. The tempering sweep is $\alpha\in\{0.05,0.1,0.25,0.5,1\}$ at $1{,}000$ walkers per population. The population sweep changes the physical and auxiliary walker counts together through $1,10,100,1000,10000$ at $\alpha=0.25$ and fixed integration and update counts, so a larger population also means proportionally more sampling work. Errors here are Boltzmann-weighted, mean-gauge FES RMSEs of trailing parameter averages, evaluated on fixed reference samples drawn from a $1200\times1200$ inverse-CDF discretization of the M\"uller--Brown density (\cref{tab:mb-sweeps}). Any $\alpha\in[0.05,0.5]$ reaches $0.09$--$0.11\,k_BT$ against $1.28\,k_BT$ for a flat surface, while $\alpha=1$ remains at $0.60\,k_BT$. The error falls from $1.95\,k_BT$ with one walker to $0.070\,k_BT$ with $10^4$.

\begin{table}[t]
\centering
\caption{M\"uller--Brown errors at $t=320$: Boltzmann-weighted FES RMSE in $k_BT$ of the parameter average, mean $\pm$ sample SD over four independent runs, against the tempering exponent at 1000 walkers per population and against the walker count at $\alpha=0.25$. A flat surface scores $1.281$.}
\label{tab:mb-sweeps}
\small
\begin{tabular}{lrlr}
\toprule
$\alpha$ & RMSE & Walkers & RMSE\\
\midrule
$0.05$ & $0.109\pm0.018$ & $1$ & $1.948\pm0.749$\\
$0.1$ & $0.098\pm0.026$ & $10$ & $0.377\pm0.160$\\
$0.25$ & $0.104\pm0.024$ & $100$ & $0.169\pm0.082$\\
$0.5$ & $0.093\pm0.029$ & $1000$ & $0.104\pm0.024$\\
$1$ & $0.602\pm0.494$ & $10000$ & $0.070\pm0.017$\\
\bottomrule
\end{tabular}
\end{table}

\subsection{Four-well dimensional scaling}\label{app:fourwell-scaling}
\paragraph{Landscape.}
The reference landscape has four symmetric Gaussian components whose centers lie in a fixed two-dimensional plane involving all $d$ coordinates. Up to an additive constant, its thermal-unit form is
\[
 \frac{G_\star(s)}{k_BT}=-\log\sum_{a,b\in\{-1,1\}}
 \exp\!\left[-\frac{\|s-\delta(a v_1+b v_2)\|^2}{2\sigma_w^2}\right],
 \qquad \delta=1.6,\quad\sigma_w=0.3.
\]
Here $v_1=(1,\ldots,1)/\sqrt d$, and $v_2$ is obtained by orthogonalizing the alternating-sign vector against $v_1$ and normalizing it. The remaining directions are harmonic, with dimensionless curvature $1/\sigma_w^2$. The adjacent-well barriers are approximately $13.5\,k_BT$. The CV map is the identity, and the dynamics use reflecting boundaries on $[-3,3]^d$. The two-dimensional core is identical at every $d$. The reference contains $40{,}000$ i.i.d. draws from the analytic teacher Boltzmann distribution, and errors use all $d$ coordinates.

\paragraph{Protocol.}
$d\in\{2,5,10,25,50,100\}$. The same three-hidden-layer network of width 1024 is used at every $d$. A separate perfect-information regression of the exact free energy into this network (i.i.d. Boltzmann samples, exact energies, $160{,}000$ Adam updates of batch $1{,}000$) establishes that it can represent the four-well plane at $d=100$ (widths 128, 256 and 512 do not learn the $d=100$ plane, while width 1024 reaches a plane error of $0.03$--$0.13\,k_BT$ against $1.29$ for a flat surface at every $d$). The primary arms use $1{,}000$ physical and $1{,}000$ auxiliary walkers in total, both started in the deepest well, $160{,}000$ overdamped steps at $\Delta t=0.005$ (reduced time $800$), Adam at $3\times10^{-5}$, a bias-corrected parameter EMA with time constant $2{,}000$ steps for the convergence readout and a trailing-half parameter average for the endpoint. Arms: single levels at $\alpha=1$, $0.5$, $0.25$, $0.1$, $0.05$, and four-level ladders ($\alpha=0.5$, $0.25$, $0.1$, $0.05$; $250$ walkers per level and population) with and without symmetric exchange of physical and auxiliary walkers. As a control on population rather than temperature, the same ladders were also run with $1{,}000$ walkers per level and population. Three independent runs were specified per arm. The unbiased $d=2$ arm uses its two completed runs, and all other arms use three.

\paragraph{Readouts.}
$\mathrm{NRMSE}=\mathrm{RMSE}/\sigma$, where the RMSE is the teacher-Boltzmann-weighted full-coordinate error after removing the additive constant and $\sigma$ is the corresponding standard deviation of the reference free energy. The endpoint NRMSE uses the trailing-half SWA readout, while threshold time uses the bias-corrected EMA readout. The reduced threshold time is when the EMA NRMSE first stays below $0.25$ to the end of the run. $D_{\mathrm{KL}}(\rho\,\|\,\rho_\theta)$ uses the same teacher-Boltzmann reference samples. Endpoint summaries are means $\pm$ sample SD (\cref{tab:fourwell-scaling,tab:fourwell-population}). For threshold-time summaries, a run that never reaches the sustained threshold is capped at the budget of $800$.

\begin{table}[t]
\centering
\footnotesize
\setlength{\tabcolsep}{3pt}
\caption{Four-well dimensional scaling: mean $\pm$ sample SD of the full-coordinate NRMSE and $D_{\mathrm{KL}}(\rho\,\|\,\rho_\theta)$ at the final SWA readout. Each best single level is selected by the lowest mean of the corresponding metric among the tested levels at that dimension. All primary arms use the same total walker budget.}
\label{tab:fourwell-scaling}
\begin{tabular}{rllllll}
\toprule
 & \multicolumn{3}{c}{NRMSE} & \multicolumn{3}{c}{$D_{\mathrm{KL}}$} \\
\cmidrule(lr){2-4}\cmidrule(lr){5-7}
$d$ & L-RE & Best single ($\alpha$) & L-NRE & L-RE & Best single ($\alpha$) & L-NRE \\
\midrule
$2$ & $0.36\pm0.20$ & $0.28\pm0.10\;(0.05)$ & $0.38\pm0.11$ & $0.07\pm0.07$ & $0.05\pm0.03\;(0.05)$ & $0.07\pm0.03$ \\
$5$ & $0.42\pm0.10$ & $0.26\pm0.09\;(0.5)$ & $0.41\pm0.31$ & $0.18\pm0.07$ & $0.08\pm0.05\;(0.5)$ & $0.22\pm0.27$ \\
$10$ & $0.23\pm0.07$ & $0.16\pm0.04\;(0.25)$ & $0.22\pm0.06$ & $0.13\pm0.07$ & $0.06\pm0.02\;(0.25)$ & $0.12\pm0.06$ \\
$25$ & $0.22\pm0.06$ & $0.20\pm0.03\;(0.5)$ & $0.16\pm0.01$ & $0.27\pm0.13$ & $0.24\pm0.08\;(0.5)$ & $0.18\pm0.01$ \\
$50$ & $0.19\pm0.01$ & $0.24\pm0.04\;(0.5)$ & $0.17\pm0.01$ & $0.41\pm0.04$ & $0.60\pm0.09\;(0.5)$ & $0.36\pm0.04$ \\
$100$ & $0.32\pm0.21$ & $0.42\pm0.05\;(0.5)$ & $0.40\pm0.19$ & $2.30\pm2.40$ & $2.40\pm0.23\;(0.5)$ & $3.49\pm2.41$ \\
\bottomrule
\end{tabular}
\end{table}

\begin{table}[t]
\centering
\small
\caption{Larger-population controls for the four-well landscape: mean $\pm$ sample SD with $1{,}000$ walkers per level and population, four times the primary ladder budget.}
\label{tab:fourwell-population}
\begin{tabular}{rllll}
\toprule
 & \multicolumn{2}{c}{NRMSE} & \multicolumn{2}{c}{$D_{\mathrm{KL}}$} \\
\cmidrule(lr){2-3}\cmidrule(lr){4-5}
$d$ & L-RE & L-NRE & L-RE & L-NRE \\
\midrule
$2$ & $0.14\pm0.03$ & $0.18\pm0.06$ & $0.010\pm0.004$ & $0.02\pm0.01$ \\
$5$ & $0.26\pm0.02$ & $0.24\pm0.04$ & $0.08\pm0.01$ & $0.07\pm0.02$ \\
$10$ & $0.27\pm0.03$ & $0.26\pm0.02$ & $0.16\pm0.03$ & $0.17\pm0.03$ \\
$25$ & $0.25\pm0.05$ & $0.20\pm0.05$ & $0.40\pm0.20$ & $0.24\pm0.13$ \\
$50$ & $0.20\pm0.02$ & $0.179\pm0.001$ & $0.47\pm0.09$ & $0.39\pm0.01$ \\
$100$ & $0.39\pm0.20$ & $0.33\pm0.11$ & $2.95\pm2.16$ & $2.07\pm0.96$ \\
\bottomrule
\end{tabular}
\end{table}

\paragraph{Observations.}
The same exchange ladder is within a factor of $1.62$ of the best tested single level on mean NRMSE across all dimensions. Its mean NRMSE ranges from $0.19$ to $0.42$, with lower mean error than the best single level at $d=50$ and $100$. At $d=100$, the primary exchange and no-exchange ladders reach $0.32\pm0.21$ and $0.40\pm0.19$, respectively, while the best single level reaches $0.42\pm0.05$. The larger-population controls reach $0.39\pm0.20$ with exchange and $0.33\pm0.11$ without (\cref{tab:fourwell-population}). These descriptive comparisons support one fixed ladder across the tested dimensions. They do not establish uniform superiority or an exchange advantage at every dimension.

\section{Connections to VES, PS-CD, and R\'enyi objectives}\label{app:related}

The following identities connect the VESPER objective to well-tempered VES, pseudo-spherical contrastive divergence, and a R\'enyi variational principle.

\subsection{Well-tempered VES and an explicit shared-energy objective}\label{app:ves-correspondence}

Let $p$ be the physical CV density and $t$ the VES target. The fixed-target functional of \citet{Valsson_2014} is
\[
 \Omega[V;t]=\beta^{-1}\log\mathbb E_p e^{-\beta V}
                  +\int t(s)V(s)\,\dd s.
\]
Differentiating with $t$ held fixed gives
\[
 \nabla_\theta\Omega[V_\theta;t]
 =-\mathbb E_{\rm biased}\nabla_\theta V_\theta
  +\mathbb E_t\nabla_\theta V_\theta.
\]
Set $V_\theta=-(1-\alpha)G_\theta$ and $t_\theta=P_{G_\theta}$, with $0<\alpha<1$. The biased physical law is then $\nu_{G_\theta}$, and substitution yields
\[
 \left.\nabla_\theta\Omega[V_\theta;t]\right|_{t=t_\theta}
 =(1-\alpha)\left\{
 \mathbb E_{\nu_{G_\theta}}\nabla_\theta G_\theta\circ s
 -\mathbb E_{P_{G_\theta}}\nabla_\theta G_\theta\right\}
 =(1-\alpha)\nabla_\theta\mathcal L_{\hat\beta}[G_\theta].
\]
This proves \eqref{eq:ves-correspondence} for every admissible model for which differentiation under the integrals is valid. The factor $1-\alpha$ can be absorbed into the step size for ordinary gradient descent with the same parameterization and sample estimate.

VESPER thus gives the instantaneous self-consistent update an explicit scalar objective, convex in the free energy. This point requires holding the VES target fixed during differentiation. Taking the total derivative of $\Omega[V_\theta;t_\theta]$ also differentiates the target and adds
\[
 \int V_\theta(s)\nabla_\theta t_\theta(s)\,\dd s
 =\alpha\beta(1-\alpha)
     \operatorname{Cov}_{t_\theta}(G_\theta,\nabla_\theta G_\theta),
\]
where we used $\nabla_\theta t_\theta=-\alpha\beta t_\theta(\nabla_\theta G_\theta-\mathbb E_{t_\theta}\nabla_\theta G_\theta)$. In contrast, differentiating $\mathcal L_{\hat\beta}$ directly gives the desired two-population update, with its minimum and KL control established in \cref{thm:consistency,prop:kl-control}.

The distinction also appears in how the target evolves. Well-tempered VES sets $t\propto p^\alpha$ \citep{valsson2015welltempered}. In neural VES, the free-energy relation $\widehat F=-V-\beta^{-1}\log t$ and the previous target $t_{n-1}\propto e^{-\alpha\beta\widehat F_{n-1}}$ give the recursive update \citep{bonatiNeuralNetworksbasedVariationally2019a}
\begin{equation}\label{eq:neural-ves-target-recursion}
 \widehat F_n=-V_n+\alpha\widehat F_{n-1}+C_n,
 \qquad t_n\propto e^{-\alpha\beta\widehat F_n}.
\end{equation}
At a fixed point, $(1-\alpha)\widehat F=-V+C$, so $V=-(1-\alpha)G$ gives $\widehat F=G+\mathrm{const}$. During learning, VESPER instead defines the target directly from the current $G_\theta$ and represents it with persistent auxiliary walkers. One model therefore determines both the molecular bias and the auxiliary potential, without a separate recursion for the free-energy estimate. The walker laws track these changing targets through the coupled dynamics analyzed in \cref{app:convergence}.

This formulation extends directly to the combined objective: every $\mathcal L_{\hat\beta_k}$ has the same physical minimizer, so all levels train one $G_\theta$ through their native gradients (\cref{thm:ladder}). In VES notation, the interior-level update is reproduced by weighting the corresponding fixed-target gradients by $c_k/(1-\alpha_k)$. The rescaled gradient has the finite physical-endpoint limit
\[
 \lim_{\alpha\uparrow1}\frac{1}{1-\alpha}
 \left.\nabla_\theta\Omega[-(1-\alpha)G_\theta;t]\right|_{t=P_{G_\theta}}
 =\mathbb E_p\nabla_\theta G_\theta
  -\mathbb E_{P_{G_\theta,\beta}}\nabla_\theta G_\theta,
\]
whenever the two expectations converge. VESPER includes this endpoint through \eqref{eq:VESPER_objective:lim}.

\subsection{PS-CD and direct sampling of the two tilted populations}

With $g=\beta G$, the dimensionless objective $\ell_\alpha$ is given by \eqref{eq:dimensionless-objective}. Let $q_g=e^{-g}/Z_g$ be the model density relative to a reference measure $r$, with $Z_g<\infty$. Substituting $\log q_g=-g-\log Z_g$ and $\gamma=\alpha-1$ gives
\begin{equation}\label{eq:pscd-correspondence}
 \ell_\alpha[g]
 =\frac1{1+\gamma}\log\int q_g^{1+\gamma}\,\dd r
     -\frac1\gamma\log\int q_g^\gamma\,\dd p
 =\mathcal L^{\rm PS}_\gamma[g].
\end{equation}
The first term contributes $-\log Z_g$ and the second $+\log Z_g$, so the normalizer cancels. This is the PS-CD objective \citep{yu2021pseudospherical}, including its negative-$\gamma$ extension, whenever both integrals are finite. For VESPER, $0<\alpha<1$ gives $\gamma=\alpha-1\in(-1,0)$. The physical free-energy minimizer follows from \cref{thm:consistency}.

For the Lebesgue reference, write $q_\theta=q_{G_\theta}=e^{-\beta G_\theta}/Z_{G_\theta}$. The two populations in the gradient are exactly the laws used in the main text:
\begin{equation}\label{eq:identity-ebm-pair}
 \begin{aligned}
 P_{G_\theta}(s)
 &=\frac{q_\theta(s)^\alpha}{\int q_\theta(u)^\alpha\,\dd u},\\
 Q_{G_\theta}(s)
 &=\frac{p(s)q_\theta(s)^{\alpha-1}}
         {\int p(u)q_\theta(u)^{\alpha-1}\,\dd u}.
 \end{aligned}
\end{equation}
Differentiating the two log integrals gives $\nabla_\theta\ell_\alpha[g_\theta]=\mathbb E_{Q_{G_\theta}}\nabla_\theta g_\theta-\mathbb E_{P_{G_\theta}}\nabla_\theta g_\theta$, recovering \eqref{eq:physical-gradient} after division by $\beta$. At the physical endpoint, $P_{G_\theta}=q_\theta$ and $Q_{G_\theta}=p$, recovering the maximum-likelihood gradient.

The computational difference is how these expectations are obtained. PS-CD uses a fixed dataset for the positive term and reweights samples from $q_\theta$ to estimate the tempered model expectation \citep{yu2021pseudospherical}. VESPER directly evolves auxiliary walkers in $\alpha G_\theta$, removing that temperature-change importance weighting from the fresh-sample gradient. In molecular applications, it also generates $Q_{G_\theta}$ online through the biased microscopic law $\nu_{G_\theta}$. This makes the same variational principle usable when equilibrium configurations are not available in advance: sampling supplies new configurations while their gradients update the free-energy model (\cref{app:proof-dynamics}).

\subsection{R\'enyi variational identity and its sampling realization}

For a probability reference $r$ and $0<\alpha<1$, set $h=-g$ in \eqref{eq:dimensionless-objective}. Then
\[
 -\ell_\alpha[g]
 =\frac1{\alpha-1}\log\int e^{(\alpha-1)h}\,\dd p
  -\frac1\alpha\log\int e^{\alpha h}\,\dd r,
\]
which is the variational expression of \citet[Theorem~3.1]{birrell2021renyi}. With $f=\dd p/\dd r>0$ $r$-almost everywhere and $M(\alpha)=\int f^\alpha\,\dd r$, \cref{prop:reference-minimizer} gives $g_\star=-\log f+C$. Substituting this minimizer into the two integrals yields
\[
 \inf_g\ell_\alpha[g]
 =\frac{\log M(\alpha)}{\alpha(1-\alpha)}
 =-\frac1\alpha D_\alpha(p\|r),
 \qquad D_\alpha(p\|r)=\frac{\log M(\alpha)}{\alpha-1}.
\]
The factor $1/\alpha$ accounts for the divergence convention in that reference. If $f$ vanishes on a set of positive $r$-measure, the infimum need not be attained by a finite free energy.

The value of this identity for VESPER is that its optimizer is also the quantity needed to drive sampling. For $r=\dd s$, $g_\star=\beta G_\star$ up to a constant. VESPER uses $G_\theta$ to bias microscopic dynamics and to evolve the auxiliary walkers, obtaining both gradient expectations without evaluating the unknown CV density $p$ or first collecting equilibrium data (\cref{app:proof-dynamics}). The KL decomposition in \cref{prop:kl-control} connects the objective excess to the discrepancy of $P_G$ and $Q_G$ from their common tempered target, and \cref{thm:ladder} lets several orders train the same physical free energy.

For a reference of finite mass $R$, normalization to $r/R$ subtracts $(\log R)/\alpha$ from $\ell_\alpha$ and leaves its gradients unchanged. On infinite-volume CV domains we instead use the exponential integrals directly, with the integrability conditions of \cref{app:proof-convexity,app:reference-measure}. This preserves the same free-energy minimizer without introducing an artificial uniform probability reference.

\section{Algorithm and implementation details}\label{app:extra}

VESPER is implemented in PySAGES \citep{zubietaricoPySAGESFlexibleAdvanced2024}. We alternate molecular and auxiliary sampling with parameter updates, retaining the auxiliary population between epochs. Neural models provide both parameter derivatives for learning and spatial derivatives for biasing. When several parameter updates reuse one batch, within-population importance weights account for the change from its generating model. The pseudocode and weight formulas are given below. Model architectures, integration parameters, optimizer settings, exchange intervals, and readouts are specified for each experiment in \cref{app:physics}.

\Cref{alg:vesper} describes physical sampling with an energy-unit model $G_\theta$ and level ratios $\alpha_k=\hat\beta_k/\beta$. Each epoch generates physical and auxiliary samples, optionally exchanges configurations (predominantly physical configurations in the reported ADP and ATP runs, \cref{app:physics}), and updates the shared model.

\begin{algorithm}[h]
\caption{Physical VESPER ladder}
\label{alg:vesper}
\begin{algorithmic}[1]
\STATE \textbf{Input:} potential $U$, CV map $s$, inverse temperature $\beta$; levels $0\le\hat\beta_1<\cdots<\hat\beta_K\le\beta$; weights $c_k>0$, $\sum_kc_k=1$; model $G_\theta$; sampler, optimizer, exchange, and readout settings.
\STATE Initialize $\theta$ and the physical and auxiliary populations on every level.
\REPEAT
 \STATE Set $\theta_0\leftarrow\theta$. Advance physical walkers in $U-(1-\alpha_k)G_{\theta_0}\circ s$ and auxiliary walkers in $\alpha_kG_{\theta_0}$, at inverse temperature $\beta$.
 \STATE If exchange is enabled, attempt adjacent-level swaps using \eqref{eq:main-exchange}.
 \STATE Save batches $\{x_k^j\}_{j=1}^{n_k}$ and $\{s_k^i\}_{i=1}^{N_k}$, together with their generating model $G_{\theta_0}$.
 \FOR{each parameter update allowed by the protocol}
  \FOR{$k=1,\ldots,K$}
   \IF{sample reweighting is enabled}
    \STATE $\Delta G\leftarrow G_\theta-G_{\theta_0}$.
    \STATE $w_j^{(x,k)}\leftarrow e^{(\beta-\hat\beta_k)\Delta G(s(x_k^j))}$,
           $w_i^{(s,k)}\leftarrow e^{-\hat\beta_k\Delta G(s_k^i)}$.
    \STATE Normalize the weights separately in each pool.
   \ELSE
    \STATE $\bar w_j^{(x,k)}\leftarrow1/n_k$, $\bar w_i^{(s,k)}\leftarrow1/N_k$.
   \ENDIF
   \STATE $h_k\leftarrow\sum_j\bar w_j^{(x,k)}\nabla_\theta G_\theta(s(x_k^j))
                    -\sum_i\bar w_i^{(s,k)}\nabla_\theta G_\theta(s_k^i)$.
  \ENDFOR
  \STATE If an enabled sample-reuse threshold is exceeded, end the inner loop and generate fresh samples.
  \STATE Update $\theta$ with the configured optimizer using $\sum_kc_kh_k$.
 \ENDFOR
 \STATE Save model checkpoints and sampling diagnostics.
\UNTIL{the stopping condition or simulation budget is reached}
\STATE \textbf{Output:} the checkpoint or averaged readout selected by the protocol.
\end{algorithmic}
\end{algorithm}

\paragraph{Sample reuse.}
The weights of \eqref{eq:si-reuse-weights} supply the two density ratios for a batch generated at $G_{\theta_0}$. Reweighted estimators also appear in \citet{carbone2023jarzynski} and \citet{invernizziRethinkingMetadynamics2020}. The molecular implementation can monitor the 95th percentile and maximum of $|\log w|$, with soft and hard thresholds $5$ and $8$, and the normalized ESS $(\sum w)^2/[n\sum w^2]$, with floor $0.2$. Activation, step reduction, and refresh rules are specified per system in \cref{app:physics-trust}.

\paragraph{Existing-data updates.}
For equilibrium observations $x_j^*\sim\varrho$, replace physical sampling and its reuse weights by \eqref{eq:main-data-weights}.

\paragraph{Checkpoint and averaged readouts.}
A trailing weight-space average produces a network with parameters
\begin{equation}\label{eq:swa}
 \bar\theta=\frac1{|W|}\sum_{k\in W}\theta_k.
\end{equation}
The checkpoint set $W$, stride, and window are specified per experiment \citep{polyakAcceleration1992,izmailovAveragingWeights2018}. Fixed-checkpoint and last-iterate readouts use their corresponding $\theta_k$.

\end{document}